\documentclass[aps,prb,twocolumn,amsmath,amssymb,eqsecnum,nofootinbib,superscriptaddress]{revtex4}
\usepackage{graphicx}
\usepackage{dcolumn}
\usepackage{bm}

\usepackage[colorlinks=true, urlcolor=violet, linkcolor=blue, citecolor=red, hyperindex=true, linktocpage=true]{hyperref}

\newcommand{\beq}{\begin{equation}}
\newcommand{\eeq}{\end{equation}}
\newcommand{\ba}{\begin{array}{ccc}}
\newcommand{\ea}{\end{array}}
\newcommand{\nn}{\nonumber}
 \renewcommand{\d}{\partial}
\def\bea{\begin{eqnarray}}
\def\eea{\end{eqnarray}}

\def\<{\langle}
\def\>{\rangle}

\def\mx{\mathrm{x}}
\def\my{\mathrm{y}}

 \usepackage{amsmath}
\usepackage{amssymb}

\begin{document}

\title{Are non-Fermi-liquids stable to Cooper pairing?}

\author{Max A. Metlitski}
\affiliation{Kavli Institute for Theoretical Physics, UC Santa Barbara, CA 93106}
\author{David F. Mross}
\affiliation{Physics Department, California Institute of Technology, Pasadena, CA  91125}
\author{Subir Sachdev}
\affiliation{Department of Physics, Harvard University, Cambridge, MA 02138}
\affiliation{Perimeter Institute of Theoretical Physics, Waterloo, Ontario, N2L 2Y5, Canada.}
\author{T. Senthil}
\affiliation{Department of Physics, Massachusetts Institute of Technology, Cambridge, Massachusetts 02139}

\date{\today}
\begin{abstract}
States of matter with a sharp Fermi-surface but no well-defined Landau
quasiparticles arise in a number of physical systems.
Examples include: ({\em i\/})~quantum critical points associated with the onset
of order in metals; ({\em ii\/})~spinon Fermi-surface (U(1) spin-liquid)
state of a Mott insulator;  ({\em iii\/})~Halperin-Lee-Read composite
fermion charge liquid state of a half-filled Landau level. In this
work, we use renormalization group techniques to investigate
possible instabilities of such non-Fermi-liquids in two spatial dimensions to  Cooper pairing. We consider the Ising-nematic
quantum critical point as an example of an ordering phase transition in a metal, and demonstrate
that the attractive interaction mediated by the order parameter fluctuations always
leads to a superconducting instability. Moreover, in the regime where our calculation is controlled,  superconductivity
 preempts the destruction of electronic quasiparticles.
On the other hand, the spinon Fermi-surface
and the Halperin-Lee-Read states are stable against Cooper pairing for a
sufficiently weak attractive short-range interaction; however, once
the strength of attraction exceeds a critical value, pairing sets in.
We describe the ensuing quantum phase transition between ({\em i\/})
$U(1)$ and  $Z_2$ spin-liquid states; ({\em ii\/})  Halperin-Lee-Read and
Moore-Read states. 
\end{abstract}

\maketitle

\section{Introduction}

It is well-known that ordinary metals described by Fermi-liquid (FL) theory are unstable to an arbitrarily weak
attractive interaction in the BCS channel, which leads to Cooper pairing of electrons and drives the system into a superconducting phase. 
The purpose of the present paper is to examine the stability of certain {\it non}-Fermi-liquid (nFL) states in two dimensions to  Cooper pairing. We  study systems where
the non-Fermi-liquid behavior arises as a result of the interaction of a gapless bosonic mode with fermions in the vicinity of the Fermi-surface (FS). 
Specific examples we analyze are described in the following subsections.

\subsection{Quantum critical points in metals}  

\begin{figure}[h]
\begin{center}
\includegraphics[width=3.3in]{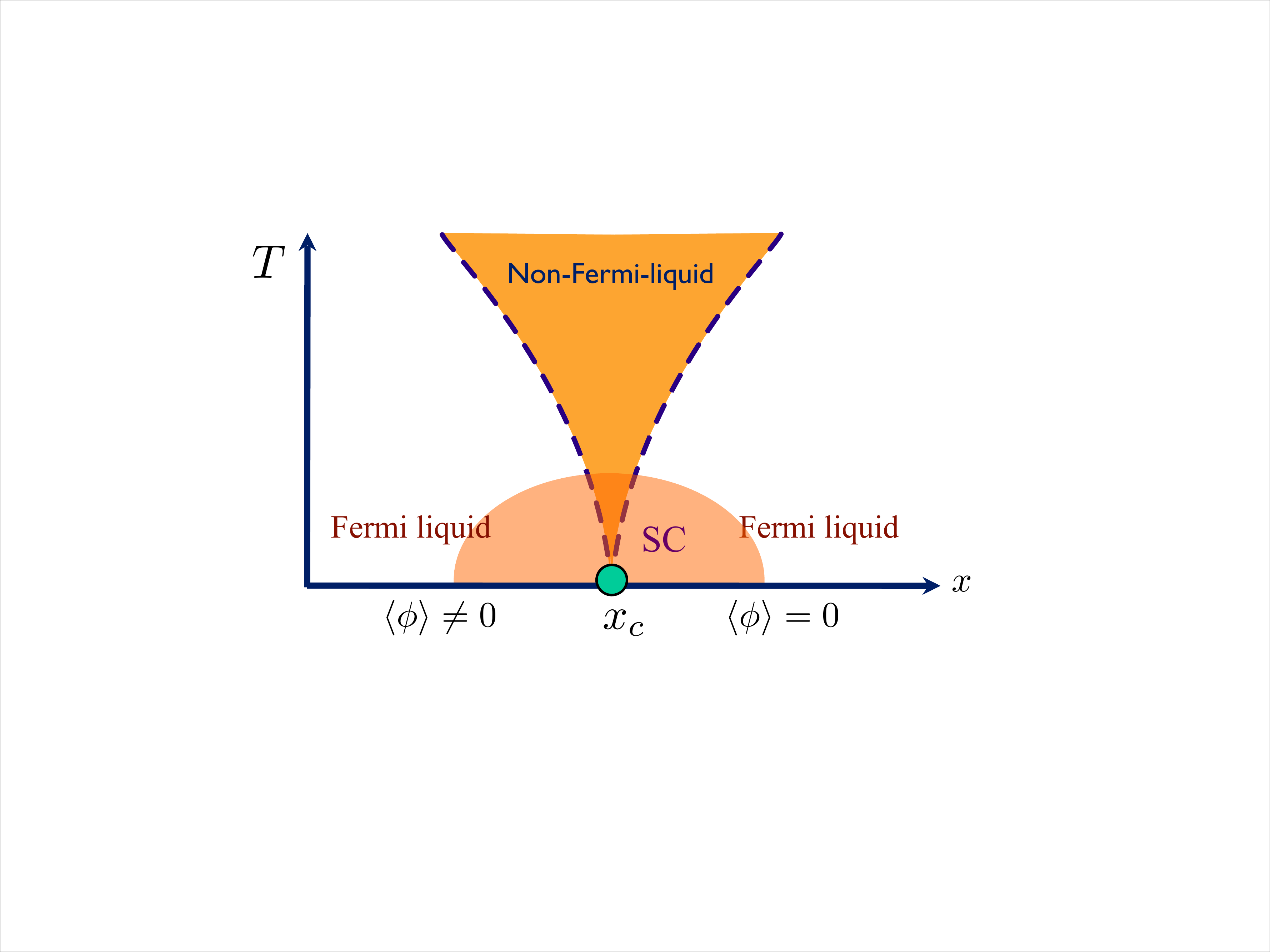}\\
\includegraphics[width=3.3in]{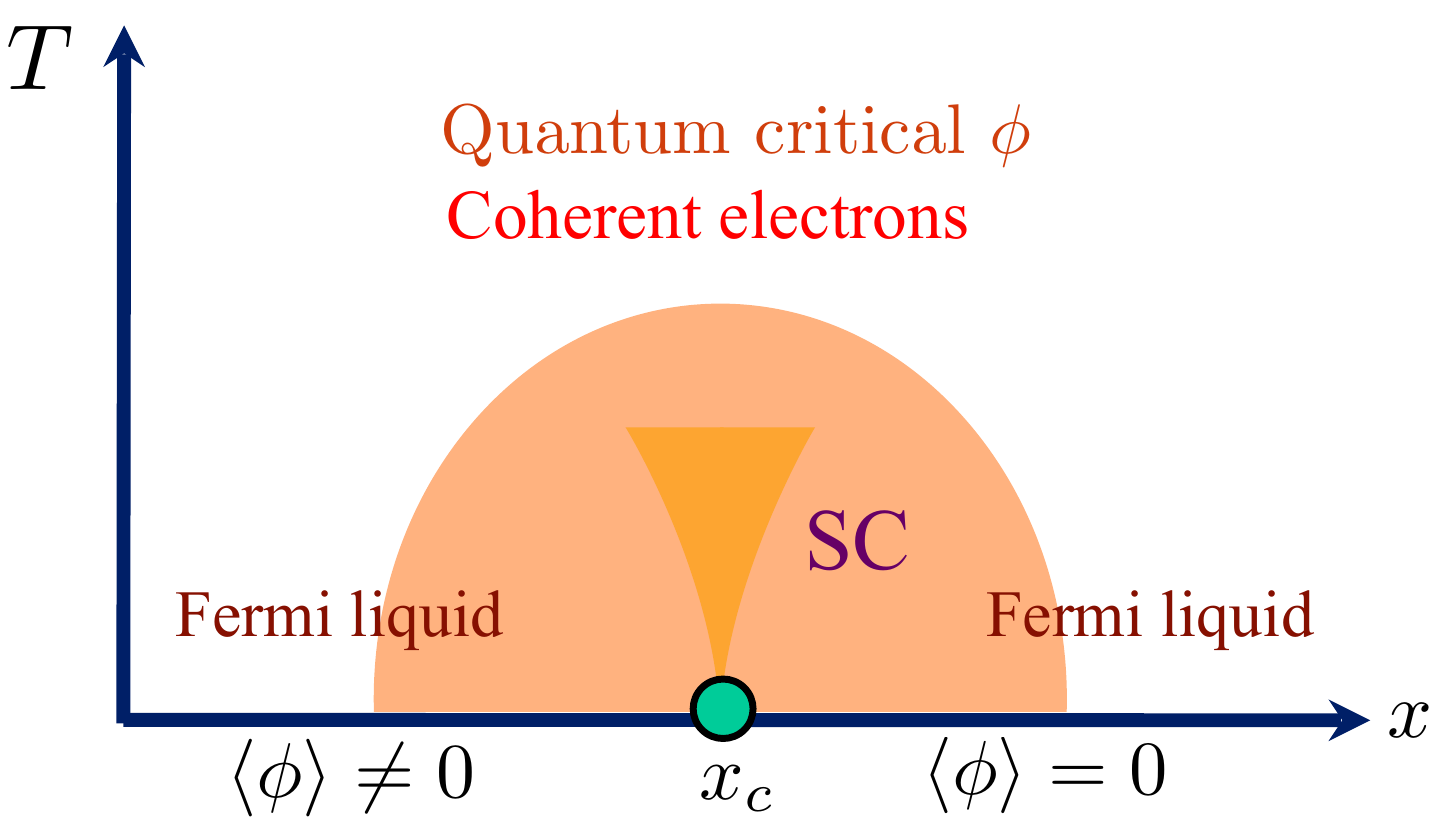}
\caption{Top: Conventional phase diagram of a quantum critical point (QCP) associated with an order parameter $\phi$, with a superconducting dome (SC)
partially overlapping the quantum critical region of the `bare' QCP of a metal.
Bottom: the phase diagram obtained in the present paper, with the SC dome fully overlapping
the incipient regime of incoherent fermionic quasiparticles, while the quantum critical $\phi$ fluctuations
survive into higher temperatures in the normal state.   
}
\label{fig:nFLphase}
\end{center}
\end{figure}

Many correlated metals appear to possess quantum critical points (QCPs) with fascinating properties.\cite{LohneysenReview,TailleferReview,MatsudaReview} Frequently, there is a striking breakdown of Fermi liquid theory in the vicinity of the QCP. Equally strikingly superconductivity is  often but not always  strengthened near the QCP. Indeed, a fairly common phase diagram (see Fig.~\ref{fig:nFLphase}, top),  shared for instance by cuprate, pnictide and certain heavy-fermion materials, has a superconducting dome around the putative `metallic' QCP with `optimal' transition temperature $T_c$ right at the QCP.  On the other hand, there are prominent quantum critical heavy electron metals such as CeCu$_{6-x}$Au$_x$ and YbRh$_2$Si$_2$ where superconductivity does not appear down to very low temperatures.\cite{LohneysenReview} It thus appears that superconductivity is enhanced at some but not all quantum critical points in metals. Despite this there is currently limited understanding of the interplay between the quantum criticality, the non-Fermi liquid `normal' state, and the possible superconductivity. Clearly, a theory of the relationship of  superconductivity and quantum criticality has to accommodate the absence of superconductivity at some and enhancement at other quantum critical points. 


It is important right away to recognize that there are two fundamentally distinct classes of quantum criticality in metallic systems. They are distinguished by the fate of the electron Fermi surface as the metal undergoes the quantum phase transition. In one class, the electron Fermi surface evolves continuously through the critical point but is distorted in some way. These QCPs are typically associated primarily with the onset of a broken symmetry characterized by a Landau order parameter in a metal. Examples include  the onset of ferromagnetism or antiferromagnetism in a paramagnetic metal. The proper  theoretical framework to describe such a phase transition is through coupling the low energy electronic degrees of freedom at the Fermi surface to fluctuations of the Landau order parameter.\cite{Hertz}  An alternate class of quantum phase transitions involves a more violent transformation of the electronic structure where the electron Fermi surface (or a sheet of it) disappears completely on crossing the critical point.\cite{coleman} Surprisingly, such a discontinuous evolution of the electron Fermi surface can happen through a continuous phase transition.  Examples include the so-called Kondo breakdown transition in Kondo lattices\cite{ffl,kbpepinetal} and continuous Mott metal-insulator transitions in two\cite{mottcrit} or 
three dimensions.\cite{mottcrit3d}  There is currently only one known theoretical framework that yields such a phase transition: 
this is based on slave-particle methods and inevitably leads to a description in terms of fractionalized slave-particles coupled to fluctuating emergent gauge fields.

In this paper we will consider examples of both kinds of quantum critical points as case studies for the relationship between quantum criticality, superconductivity, and non-Fermi liquid physics. In the example studied of the first class, where the Fermi surface is distorted through the development of a broken symmetry, we show that superconductivity is strongly enhanced near the critical point. We suggest that this may be more generally true: order parameter fluctuations enhance superconductivity.  In the example studied of the second class where the entire electron Fermi surface is annihilated, we argue that superconductivity is suppressed. This dichotomy may explain the phenomenology described above where some but not all QCPs show an enhancement of superconductivity. 

We begin with QCPs associated with the onset of a symmetry breaking order. Strong fluctuations of the order parameter present at the QCP tend to decohere the electronic quasiparticles: as the system is tuned to the critical point, the residue $Z$ and the Fermi-velocity $v_F$ of quasiparticles approach zero.  A common feature of such QCPs is that there exists some pairing channel in which the order parameter fluctuations mediate attraction. The strength of the attraction increases as one approaches the QCP, yet the same order parameter fluctuations, which provide the pairing glue, also destroy the very quasiparticles that are trying to pair. The central question is which of these two competing effects wins. In particular, is such a QCP in a metal inherently unstable to superconductivity, as  empirical observations suggest?\cite{footnoteZaanen}

In the present paper we address the above question for the class of metallic QCPs, where the order parameter carries a wave-vector $\vec{Q} = 0$ (for recent progress on the $\vec{Q} \neq 0$ case, see Refs. \onlinecite{maxsdw,BergQMC}).
The most familiar example of such a phase transition is the Stoner instability associated with the development of ferromagnetic order.  Modern developments show that due to fluctuation effects the Stoner transition is likely modified at low temperature and becomes first order (or develops an intermediate spiral ordered phase).\cite{BKVchiFL,BKVFO,ChubukovPepinRech2004,ChubukovPepinRech2006, MaslovChubukovFerro}  A different example which does not suffer from these complications\cite{ChubukovPepinRech2006,MS}  (see, however, footnote \onlinecite{NematicFO}) is the transition associated with the onset of Ising-nematic order, characterized by spontaneous breaking of a four-fold rotational symmetry of the lattice to a two-fold subgroup.\cite{NematicReview, YK00, HM00, OKF01,MRA03, KKC03, YOM05, LFBFO06,LF07, JSKM08,MC09,MS,Mross, metznervert, bartosch, dalidovich, sur} The order parameter in this case is just a real Ising field $\phi(x)$. Growing evidence for such order has been found in a number of physical systems including cuprate,\cite{ando02, hinkov08a,kohsaka07, taill10b, lawlerExp,mesaros,kohsaka12} pnictide\cite{pnic3,pnic4,pnic5,pnicorbARP,pnic6,pnic7, pnic8,pnicm66} and ruthenate\cite{borzi07} materials. From a theoretical viewpoint, the Ising-nematic QCP is perhaps one of the simplest phase transitions in metals.
It, thus, provides a convenient setting for studying the interplay between quantum criticality, nFL and pairing physics.\cite{Vafek} 

We perform a systematic renormalization group (RG) analysis of the Ising-nematic QCP. Our approach utilizes an idea introduced by D.~T.~Son in his study of quark pairing by the color gauge field in dense baryonic matter.\cite{sonqcd} We combine the conventional Fermi-liquid RG treatment of Refs.~\onlinecite{Shankar, PolchinskiTASI} with the so-called ``two-patch" scaling approach of Refs.~\onlinecite{AIM, PolchinskiSpinon, SSLeeN, MS}. Analytical control is gained through the $\epsilon$-expansion introduced in Ref.~\onlinecite{nayak} and its subsequent large-$N$ improvement.\cite{Mross} We find that the Ising-nematic QCP is always unstable to superconductivity.  In particular,  attractive pairing interaction mediated by the order parameter fluctuations dominates over other residual short range interactions (even if they are repulsive) and drives a pairing instability as the QCP is approached. However, the residual short range interactions determine the angular momentum/spin channel in which the pairing instability occurs; as a result, the pairing symmetry is non-universal. The usual weak coupling BCS formula, $ T_c \sim \exp(-1/|V|)$, relating the superconducting $T_c$ to the strength of the short-range interaction $ V$ clearly does not hold in the vicinity of the QCP.  Rather the superconductivity is strongly enhanced, and $T_c$ at the QCP scales in a power-law manner with the coupling between order-parameter fluctuations and the electrons.  Thus, in this example we clearly demonstrate the importance of quantum criticality in optimizing the superconducting $T_c$. 
Moreover, in the regime where our calculation is controlled (small $\epsilon$), the energy scale at which superconductivity sets in is parametrically larger than the energy scale at which electronic quasiparticles are destroyed. Thus, the superconducting instability is so strong that it preempts the nFL physics (see the bottom figure in Fig.~\ref{fig:nFLphase}).
The above results of our RG analysis are in exact agreement with a direct solution of Eliashberg-like integral equations, as is shown elsewhere by one 
of us.\cite{MMpaper,Yuzbashyan}

Next, we proceed to the class of QCPs associated with annihilation of the Fermi-surface. We take as an example the Mott transition from a Fermi-liquid  to an insulating spin-liquid with a spinon Fermi-surface.\cite{mottcrit} Applying the RG procedure described above, we show that spinon pairing is suppressed both in the spinon Fermi-surface phase and at the Mott transition itself. As a result, the Mott transition and the FL phase in its vicinity will be stable to superconductivity. We expect similar conclusions to hold for the Kondo breakdown transition in the Kondo lattice. First, however, we review the construction and properties of the spinon Fermi-surface phase.


\subsection{Spinon Fermi-surface phase}
 
The spinon Fermi-surface phase is an exotic Mott-insulating spin-liquid with emergent spin-$1/2$ fermionic {\it spinon} excitations, $f_\alpha(x)$, $\alpha = \uparrow, \downarrow$.\cite{LNWReview}  The spinon dispersion is such that they form a Fermi-surface.  This phase may be accessed in the slave-particle (parton) treatment  of a lattice spin model, where electron spin operators $\vec{S}_i$ are represented as $\vec{S}_i = \frac12 f^{\dagger}_{i \alpha} \vec{\sigma}_{\alpha \beta} f_{i \beta}$, subject to the local constraint, $f^{\dagger}_{i \alpha} f_{i \alpha} = 1$.  While the spinons are neutral under the physical electromagnetic field, they carry a charge under an emergent $U(1)$ gauge field $a_{\mu}$, hence this phase is also often referred to as a $U(1)$ spin-liquid. An effective Lagrangian of the spinon FS phase may be written as,
\beq L_f = f^{\dagger}_\alpha [\d_\tau - i a_\tau +\epsilon(-i \nabla - \vec{a})]  f_\alpha + \frac{1}{2 g^2} (\epsilon_{\mu \nu \lambda} \d_{\nu} a_{\lambda})^2 + \ldots \label{eq:Lfintro}\eeq
where $\epsilon(\vec{k})$ is the spinon dispersion and the ellipses denote additional perturbations, such as four-spinon interactions.

The spinon FS phase is expected to naturally arise in so-called ``weak" Mott insulators - ones proximate to a metal-insulator transition. 
In this situation the spinon FS state may be conveniently described within a slave particle description of an electronic Hubbard model. We write the electron operator as $c_{i \alpha} = b_i f_{i \alpha}$, where $b$ is a charge-$e$ boson with zero spin, and $f_\alpha$ is the spinon as described above. This representation introduces a $U(1)$ gauge redundancy under which $b$ carries gauge charge $-1$ and $f_\alpha$ carries gauge charge $1$.  We consider a state in which $f_\alpha$ form a Fermi surface. If in addition the boson $b$ is condensed, we obtain the usual metallic Fermi-liquid. If, however, $b$ is gapped, we obtain an electrical insulator but with a spinon Fermi-surface coupled to a fluctuating $U(1)$ gauge field. This is the spinon Fermi-surface state introduced above. We note that right at the Mott metal-insulator transition, the boson $b$ is critical while the spinon continues to form a Fermi-surface. For now, we focus on the insulating spinon Fermi-surface phase; properties of the Mott transition will be reviewed in the next section.

There is numerical evidence for the presence of the spinon FS spin-liquid phase in the triangular lattice Hubbard model in the intermediate range of $U/t$.\cite{LesikRing, LeeLee, LesikRingDMRG, Lauchli}
Moreover, it has been proposed as a candidate  for the quasi-2d triangular lattice organic insulators $\kappa-\mathrm{(BEDT-TTF)_2Cu_2(CN)_3}$ and $\mathrm{EtMe_3Sb[Pd(dmit)_2]_2}$ (abbreviated $\kappa$ET and DMIT below).\cite{LesikRing, LeeLee} These materials have an estimated spin-exchange coupling $J\sim 250$ K, yet display no magnetic order down to $20-30$ mK temperature. Moreover, these electrical insulators, surprisingly, show metallic behavior in their low temperature spin-susceptibility ($\chi_s \to {\rm const}$)\cite{Kanoda2003, DMITNMR2008, DMITchi} and specific-heat ($C/T \to {\rm const}$).\cite{kETC, DMITC} DMIT also exhibits metallic thermal transport at low temperature, $\kappa/T \to {\rm const}$,\cite{DMITTransp} while $\kappa$ET shows activated thermal transport, albeit with a rather small gap $\Delta \approx 0.46$ K.\cite{kETtransp} Both materials can be driven metallic by an application of a moderate pressure of $\sim 0.4$ GPa, with $\kappa$ET developing superconductivity below $\sim 3$ K on the high-pressure side.\cite{kETMott} At ambient pressure, $\kappa$ET displays a phase transition (or a very rapid crossover) at $6$K,\cite{kETlatt} resulting in partial  loss of low-energy excitations as evidenced by specific heat.\cite{kETC}  
 It has been suggested that this low-temperature anomaly may be due to a pairing instability of the spinon FS.\cite{Ampere, TarunNode}

Current theoretical understanding of the spinon FS phase is based on the following observations. The presence of gapless spinon excitations in the vicinity of the FS strongly affects the gauge field dynamics. The longitudinal fluctuations of the emergent electric field are Debye screened by the spinon FS and become gapped. The fluctuations of the emergent magnetic field are Landau-damped by the FS, but remain gapless. The coupling of these Landau-damped magnetic field fluctuations to spinons is expected to lead to ``non-Fermi-liquid" behavior of the spinon FS,\cite{LeeNagaosa, KimFurusaki, AIM, PolchinskiSpinon} {\em e.g.} the anomalous scaling of specific heat $C \sim T^{2/3}$.

In this paper we analyze whether the spinon FS phase is stable to BCS pairing of spinons. We first observe that the gapless fluctuations of the magnetic field mediate a long-range repulsive interaction in the BCS channel and hence are not expected to cause spinon pairing. Indeed, fluctuations of the magnetic field mediate a current-current interaction. The spinons in a BCS pair have opposite momenta and opposite currents and hence, by Ampere's law, repel. Therefore, gapless gauge field fluctuations suppress spinon pairing.\cite{FNAmperian}   However, in addition to gauge field mediated long-range interactions, short-range interactions between the spinons will generally be present. Depending on the microscopic details of the system, such short range interactions may be attractive in the BCS channel with some angular momentum and spin. If the short range attraction is sufficiently strong, we expect the spinons to pair, developing a condensate $\langle f f \rangle$ (we leave the angular/spin structure of the pair wave-function implicit for now). As in an ordinary superconductor, the spinon excitations acquire a gap, except possibly at symmetry dictated (or accidental) point nodes on the FS. The pair condensate spontaneously breaks the emergent $U(1)$ gauge symmetry down to a $Z_2$ subgroup. As a result, the gauge field becomes gapped through the Higgs mechanism. Gauge excitations now take the form of gapped vortices carrying a magnetic flux $\pi$.
Such excitations are often referred to as visons. Visons and spinons possess mutual semionic statistics. Thus, the paired phase of spinons is just a $Z_2$ spin-liquid.

We confirm the above intuitive picture with a systematic RG calculation. We show that the spinon FS phase, is, indeed, stable as long as the strength of the short-range attractive BCS interactions $|V_m|$ is smaller than a critical value $|V_c|$ for all angular momentum channels $m$ (we employ a sign convention where $V < 0$ represents an attractive interaction). However, once $|V_m| > |V_c|$ for some $m$, the spinon FS develops an instability to pairing in angular momentum channel $m$.  $V_m = V_c$, thus, marks the quantum phase transition between the $U(1)$ spin-liquid and the $Z_2$ spin-liquid. We find the phase transition to be continuous and calculate the critical exponents using the $\epsilon$-expansion of Refs.~\onlinecite{nayak,Mross}. Our findings are contrary to previous claims\cite{UbbensLee} that this phase transition is driven first order by gauge field fluctuations.  We discuss the properties of the paired phase in the vicinity of the transition. Right at the critical point 
we find (at least to the order of the $\epsilon$-expansion that we study) that most experimentally accessible properties (specific heat, uniform and finite wave-vector spin-susceptibility, spin-chirality correlations)  are not modified from those in the spinon Fermi-surface phase itself. Our findings are in exact agreement with an Eliashberg-like treatment of the problem.\cite{MMpaper}

Previously, the pairing quantum phase transition from the spinon Fermi-surface state was considered in 3 dimensions by Chung {\it et al.}\cite{sudippair} within an Eliashberg-like approximation. Our paper presents an RG analysis directly in 2 dimensions, although there are some qualitative similarities with the results of  Chung {\it et al.}\cite{sudippair} In particular, Chung {\it et al.} have also concluded that a continuous pairing transition is possible. However, we believe that some of the results of Chung {\it et al.} are not generic. In particular, Chung {\it et al.} find that pairing can only occur in angular momentum channels $m \ge 2$. In contrast, we believe that both in 2d and 3d pairing with arbitrary angular momentum can be induced by tuning the appropriate $V_m$. Furthermore, we expect the power-law onset of the pairing gap found by Chung {\it et al.} in 3d to be modified by the renormalization of spinon quasiparticle residue and Fermi-velocity. In fact, we anticipate that the precise critical properties of the pairing transition in 3d will be very similar to those of the 2d Halperin-Lee-Read phase in the presence of long-range Coulomb interactions, discussed in section \ref{sec:introHLR}.

\subsection{Mott transition from a Fermi-liquid to a spinon Fermi-surface phase}
\label{sec:introMott}
The Mott transition from a Fermi-liquid to a spinon FS phase is an example of a QCP where the entire electron FS disappears. As noted in the previous section, this transition is driven by condensation of the slave boson $b$. 
The transition may be described by the effective theory,
\beq L = L_b + L_f \label{eq:Lmott}\eeq
where the Lagrangian $L_b$ for the complex scalar field $b$ is
\beq L_b = |(\d_\tau - i a_\tau) b|^2 + v^2_b |(\nabla - i \vec{a}) b|^2 +  t |b|^2 + u |b|^4 \label{eq:Lb} \eeq
and $L_f$ is still given by Eq.~(\ref{eq:Lfintro}). Note that here we are considering a Mott transition occurring at fixed electron density. When $t$ is large and positive, the boson $b$ is gapped and can be integrated out, so the system is in the spinon FS phase. On the other hand, when $t$ is large and negative, $b$ is condensed, $\langle b \rangle \neq 0$. As a result, the gauge field $a_{\mu}$ becomes gapped via the Higgs mechanism; furthermore, the electron $c_\alpha = b f_\alpha $ and the spinon $f_\alpha$ are identified, $c_\alpha \to \langle b\rangle f_\alpha$.  Thus, the system is in the ordinary FL phase.

We now discuss the fate of the system when $t$ is tuned to a critical value $t_c$ where $b$ is gapless (for more details, see Ref.~\onlinecite{mottcrit}, whose findings we summarize here). 
If the fluctuations of the gauge field $a_{\mu}$ are ignored then the spinon and boson sectors in Eq.~(\ref{eq:Lmott}) decouple, and the boson sector undergoes a transition in the XY universality class, while the spinon sector remains a ``spectator" Fermi-liquid across the transition. Proceeding to include gauge field fluctuations, we note that the longitudinal electric field is again Debye screened by the spinon Fermi-surface and so can be ignored. The fluctuations of the magnetic field are again Landau-damped by the spinon Fermi-surface, but remain gapless. It turns out that such Landau-damped gauge fluctuations do not affect the $b$-sector of the theory, which remains decoupled from the spinon sector and continues to be described by the XY critical theory. On the other hand, the $b$-sector does affect the low energy gauge fluctuations. Integrating the gapless $b$ boson at the XY critical point out, one obtains the following effective action for the magnetic field fluctuations,
\beq S_a =  \frac12 \int d^2 \vec{x} d^2 {\vec{x}}' d\tau (\nabla \times \vec{a})(\vec{x}, \tau) \Pi(\vec{x} - \vec{x}') (\nabla \times \vec{a})(\vec{x}', \tau) \label{eq:bosonRPA} \eeq
where $\Pi(\vec{x}) = v_b \sigma/(4 \pi^2 |\vec{x}|)$, and $\sigma \approx 0.36$ is the universal conductivity of the $XY$ model.\cite{XYcond} Thus, the gauge-spinon sector of the theory is described by the action 
\beq S = \int d^2 x d \tau L_f + S_a \label{eq:SMott}\eeq
which will be the starting point for our theoretical analysis in this paper. We note that this action coincides with that of the Halperin-Lee-Read state with Coulomb interactions, discussed in section \ref{sec:introHLR}. Studying the theory (\ref{eq:SMott}), one finds that gauge field fluctuations turn the spinon FS at the Mott transition into a marginal Fermi-liquid with a specific heat $C \sim - T \log T$, which dominates the overall specific heat of the system. We remind the reader that since the physical electron $c_\alpha$ is a product of the boson $b$ and the spinon $f_\alpha$, the actual physical electron Green's function displays a ${\it strong}$ deviation from Fermi-liquid theory at the Mott transition.

One may now ask whether the spinon FS at the Mott transition is stable to BCS pairing of spinons. Before we address this question, we would like to stress that independent of whether the spinons pair, we expect no long-range superconductivity exactly at the Mott transition. After all, at the Mott transition charge degrees of freedom are on the verge of becoming localized so long-range phase coherence will be suppressed. Instead, spinon pairing should be interpreted as a {\it local} tendency of electrons to pair. Let us discuss the scenario where spinon pairing does occur at the transition. In this case, as we tune the system away from the Mott transition to $t < t_c$, a  condensate $\langle c c \rangle \sim \langle b \rangle^2  \langle f f \rangle$ appears, {\it i.e.} the compressible phase adjacent to the Mott transition is a superconductor rather than a Fermi-liquid. On the other hand, the phase with $t > t_c$, where the boson $b$ is gapped, is a $Z_2$ spin-liquid insulator as discussed in the previous section. Thus, if the spinons are paired the Mott transition occurs between a superconductor and a $Z_2$ spin-liquid insulator.\cite{SenthilFisher} As we approach the transition from the superconducting side, both the superconducting $T_c$ and the superconducting condensate $\langle c c \rangle$ will vanish, however, the gap to a single electron $c_\alpha$ will remain finite across the transition. In contrast, if the spinon FS is stable against pairing then the single electron gap at the transition will vanish.

With the above remarks in mind, we now summarize the conclusions of our RG analysis. As with the spinon FS phase, we show that repulsive current-current interactions mediated by the gauge field suppress spinon pairing at the Mott transiton. As a result, as long as the strength of short-range attraction between spinons $|V_m|$ is below a critical value $|V_c|$, the spinon FS at the Mott transition is stable. We believe that in this regime a stronger statement actually holds: no spinon pairing occurs on either side of the Mott transition, in particular, no superconductivity develops in the FL phase adjacent to the Mott transition. Thus, the Mott transition is an example of a QCP in a metal, which is stable to superconductivity.

On the other hand, once $|V_m| > |V_c|$ for some $m$, the spinons at the Mott transition pair, developing a condensate $\langle f f \rangle \neq 0$. Thus, in this parameter regime the Mott transition occurs between a superconductor and a $Z_2$ spin-liquid insulator, and the single electron gap remains finite across the transition.


\subsection{Halperin-Lee-Read phase} 
\label{sec:introHLR}

 The RG formalism developed in this paper can be applied to analyze the stability of yet another exotic phase: the Halperin-Lee-Read (HLR) phase. The HLR phase is a compressible phase of the quantum Hall (QH) fluid at a filling fraction $\nu = 1/2$.\cite{HLR} It is believed to be experimentally realized by the conventional 2DEG in the lowest Landau level.\cite{WillettAP} When the Landau level is half-filled, there are two magnetic flux quanta per each electron. If one performs a transformation to composite fermions (CF) by attaching two flux quanta to each electron, the composite fermions will, on average, see no magnetic field and form a Fermi-surface. Technically, flux attachment is performed with an aid of a Chern-Simons (CS) $U(1)$ gauge field $a_{\mu}$, leading to the action 
\bea
S &=&\int d^2x d \tau (L_{f} + L_{CS}) + S_{U}, \nn \\
L_{f} &=& f^{\dagger} [\d_\tau - i a_\tau -\frac{1}{2 m}(\d_i - i a_i + i A_i)^2 ] f \\
L_{CS} &=& \frac{i}{2 (4 \pi)} \epsilon_{\mu \nu \lambda} a_{\mu} \d_{\nu} a_{\lambda} \label{eq:LCS} \\
 S_{U} &=& \frac12 \int d^2 \vec{x} d^2 \vec{x}' d\tau f^{\dagger} f(\vec{x},\tau) U(\vec{x}-\vec{x}') f^{\dagger} f(\vec{x}', \tau) \nn\\\eea
Here, $f(x)$ is the composite fermion operator, $\vec{A}$ is the vector potential for the external magnetic field and $U(\vec{x})$ is the microscopic electron-electron interaction potential. Integration over $a_\tau$ produces the constraint, 
\beq \nabla \times \vec{a} = 2 (2\pi) f^{\dagger} f \label{eq:FluxConstr}\eeq
 linking the magnetic flux density of the CS field $a_{\mu}$ to the electron density $ f^{\dagger} f$. This constraint can be used to rewrite $S_U$ in terms of $\nabla \times \vec{a}$. 

At $\nu = 1/2$, the flux of the CS gauge field $a_{\mu}$ on average cancels the external magnetic field, however, fluctuations of $a_{\mu}$ about the average flux persist. The dynamics of $a_{\mu}$ are nearly the same as in the spinon FS phase with the longitudinal electric field Debye screened and gapped, and the magnetic field Landau damped and gapless. As the electric field is gapped, the CS term in Eq.~(\ref{eq:LCS}) is irrelevant in the RG sense (more precisely, it generates a charge-current interaction of composite fermions which is suppressed in the small momentum limit compared to the current-current interaction). Therefore, the low-energy effective theory of the HLR phase is nearly identical to that of the spinon FS phase when the microscopic electron interaction $U(\vec{x})$ in the QH fluid is short ranged. 
For a power law interaction, 
\beq 
U(\vec{x}) \sim \frac{1}{|\vec{x}|^{1+\epsilon}} ,
\eeq
with $\epsilon < 1$, the electron density fluctuations and hence the gauge field fluctuations are suppressed.\cite{HLR} In fact, for $\epsilon < 0$, the composite fermion quasiparticles remain sharply defined, while for Coulomb interaction, $\epsilon = 0$, the HLR phase is believed to be a marginal Fermi-liquid with a specific heat $C \sim - T \log T$.\cite{HLR,nayak} For $\epsilon > 0$, the HLR phase is a true nFL,\cite{HLR, KimFurusaki, YBcompr, YBBoltz} with a power law specific heat $C \sim T^{2/(2+\epsilon)}$, however, the  theory is under analytical control in the limit $\epsilon \ll 1$.\cite{nayak, Mross}.

In passing, we note that the HLR phase may alternately be described within a slave particle approach that exposes the conceptual similarity to a spin-liquid Mott insulator with a spinon Fermi-surface discussed above. We represent the electron operator $c$ as a product of a charge-$e$ boson $b$ and a charge neutral fermion $f$: $c = bf$. Then the bosons are at filling factor $\nu = 1/2$ and we take them to be in the bosonic Laughlin state at that filling. Being neutral, the fermions $f$ see no magnetic field, and form a Fermi-surface.  This slave particle description introduces a $U(1)$ gauge redundancy, with $b$ and $f$ carrying opposite charges under an emergent gauge field $a_{\mu}$. The corresponding gauge constraint fixes the number density of the bosons to equal that of the $f$ fermions. Being electrically charged, the boson density is simply equal to the physical electron density. Thus, the density of $f$ fermions also equals the physical electron density. Consequently, the size of the $f$ Fermi-surface is set by the physical electron density. Since the bosonic $\nu = 1/2$ Laughlin state is gapped, we can integrate the boson degrees of freedom out, generating a Chern-Simons term (\ref{eq:LCS}) for the emergent gauge field $a_{\mu}$. Thus, the slave particle description is completely equivalent to the familiar flux-attachment picture described above.

In this paper, we address the stability of the HLR phase to BCS pairing of composite fermions. As with the spinon FS phase, the long-range current-current interaction mediated by gapless gauge field fluctuations suppresses pairing in the BCS channel. Thus, we find that the HLR phase is stable as long as the strength
 of the short-range attractive BCS interaction $|V|$ is smaller than a critical value $|V_c|$. However, once $|V| > |V_c|$, pairing of composite fermions will occur, giving rise to an incompressible QH phase with a Hall conductivity $\sigma_{xy} = 1/2$. A possible ``microscopic" source of an attractive BCS interaction is the short-distance part of the charge-current interaction mediated by the CS gauge field, which produces attraction in the $p+ip$ channel.\cite{GWW} In fact, if pairing occurs in the $p+ip$ channel, the resulting phase is just the familiar Moore-Read (MR) ``Pfaffian" state.\cite{GWW} After the pairing transition, composite fermions become gapped neutral fermion excitations of the MR phase. Gauge excitations are also gapped through the Higgs mechanism and appear in the form of vortices carrying magnetic flux $\pi$ of $a_{\mu}$, which via Eq.~(\ref{eq:FluxConstr}), translates into physical electric charge $q  =e/4$. Furthermore, these vortices support Majorana zero modes of composite fermions in their core and, therefore, can be identified with $q = e/4$ non-Abelian quasiparticles of the MR state. We find the phase transition between the HLR and the MR phases to be continuous, consistent with numerical simulations \cite{HaldaneRezayi,MollerCooper,Zlatko3b}, but contrary to previous theoretical claims.\cite{BonesteelPairBreak} We describe how the neutral fermion gap and the charge gap vanish as one approaches the QCP from the MR side, and discuss the phenomenology of the MR phase in the vicinity of the transition.

\section{Renormalization Group  analysis} 

Although  various nFL states described above arise in very different physical systems, they admit a unified theoretical treatment involving a gapless $\vec{Q} = 0$ boson interacting with the FS. We denote the boson as $\phi(x)$: it represents the order parameter in the case of the Ising-nematic QCP and the transverse component of the vector potential $\vec{a}$ in the case of the spinon FS phase, the Mott critical point, and the HLR phase. We denote the fermions (physical electrons in the Ising-nematic case, spinons in the spinon FS phase/Mott transition case and composite fermions in the HLR case) as $f_\alpha$. We take the flavor index $\alpha$ to run from $1$ to $N$. Physically, $N = 2$ (two spin flavors) for the Ising-nematic QCP and spinon FS phase/Mott transition, and $N = 1$ for the spin-polarized HLR phase.

\begin{figure}[h]
\begin{center}
\includegraphics[width=2.4in]{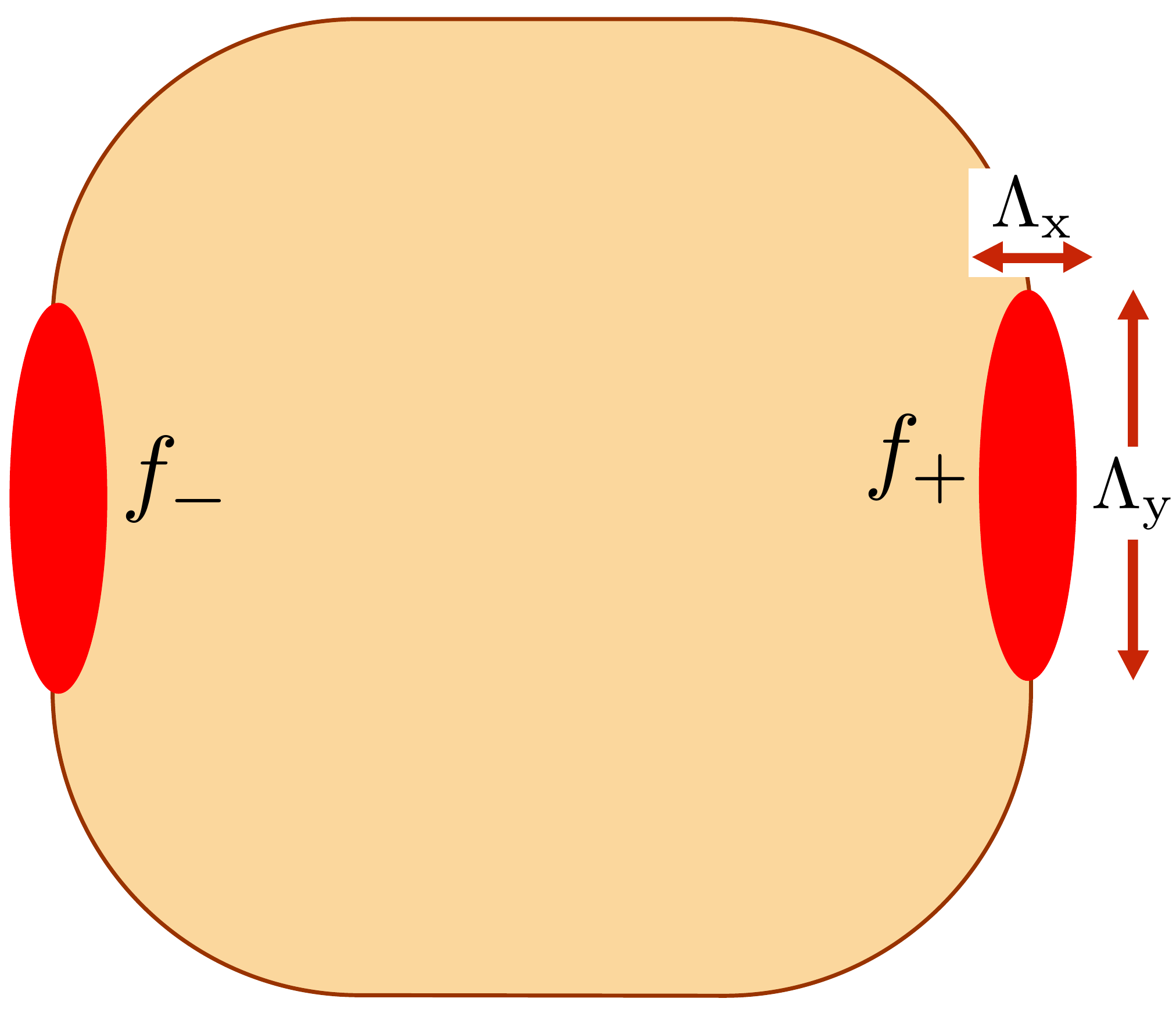}
\caption{A pair of antipodal patches, labeled by a fixed $j$, on the Fermi surface.
The values of $\Lambda_x$ and $\Lambda_y$ are constrained as in Eq.~(\ref{lambdaxy}).
}
\label{fig:2patch}
\end{center}
\end{figure}

Due to Landau-damping, boson fluctuations with wave-vector $\vec{q} \to 0$ interact most strongly with fermions in the regions of the FS to which $\vec{q}$ is nearly tangent.\cite{AIM, PolchinskiSpinon, SSLeeN}  We divide the FS into pairs of antipodal patches, labelled by an index $j$, with 
\bea
&& \mbox{width~} \Lambda_{\my} \ll k_F \nn \\
&& \mbox{and  thickness~} \Lambda_{\mx} \sim \frac{\Lambda^2_\my}{k_F} \ll \Lambda_\my \ll k_F, \label{lambdaxy}
\eea
where $k_F$ is the Fermi momentum; see Fig.~\ref{fig:2patch}. For simplicity, we assume that the Fermi-surface is connected and convex, and furthermore, that the local Fermi-surface curvature $K$ and Fermi momentum $k_F$ are comparable. 

Antipodal points $\pm \vec{k}_j$ on the FS are chosen in patch pair $j$ and directions perpendicular ($\hat{\mx}_j$) and tangent ($\hat{\my}_j$) to the FS at $\vec{k}_j$ defined. The fermion operator $f_\alpha$ is then expanded in terms of patch fields 
$f^j_{\pm, \alpha}(x)$ as, 
\beq
f_{\alpha} (x) = \sum_j (f^j_{+ \alpha}(x) e^{i \vec{k}_j \cdot \vec{x}} + f^j_{- \alpha}(x) e^{-i \vec{k}_j \cdot \vec{x}}).  
\eeq
We also define boson patch fields $\phi^j(x)$ to include only momenta nearly tangent to the FS in patch $j$: 
\beq
|q_\mx| < |q_\my| \frac{\Lambda_\my}{k_F}\,,\quad |q_\my| < \Lambda_\my. 
\eeq
The effective action $S$ describing the fermion-boson interaction then breaks up into decoupled actions for each patch pair, 
\begin{equation}
S = \sum_j S^j, 
\end{equation}
with\cite{SSLeeN, MS, Mross}
\begin{equation}
S^j = \int d^2x d\tau (L_f[f^j]  + L_{int}[f^j, \phi^j]) + S_\phi[\phi^j], 
\end{equation}
The Lagrangian densities are 
\bea L_f &=& f^{\dagger}_{+ \alpha} \left(\d_\tau + v_F (-i \d_\mx -\frac{\d^2_\my}{2 K})\right) f_{+ \alpha} \nn \\
&+& f^{\dagger}_{- \alpha}\left(\d_\tau + v_F (i \d_\mx -\frac{\d^2_\my}{2 K})\right) f_{- \alpha}  \label{eq:Lf}\\ 
L_{int} &=& v_F \phi (f^{\dagger}_{+\alpha} f_{+\alpha} + \zeta f^{\dagger}_{- \alpha} f_{- \alpha}) \label{eq:Lint}\\
S_{\phi} &=& \frac{N}{2 g^2} \int \frac{d^2 \vec{q} d \omega}{(2\pi)^3} |q_\my|^{1+\epsilon} |\phi(\vec{q}, \omega)|^2 \label{eq:Sphi} \eea
Here, we have suppressed the patch index $j$. The Fermi-surface curvature $K$, the Fermi-velocity $v_F$ and, in the case of the Ising nematic transition, the coupling constant $g^2$, will generally vary along the Fermi-surface ({\em i.e.\/} will be patch-dependent). The constant $\zeta = 1$ for the nematic QCP and $\zeta = -1$ for the spinon FS phase/Mott transition and the HLR phase. 

For general $\epsilon$, the action $S_\phi$ is non-local. For the HLR state, this term encodes the long-range microscopic electron-electron interation, $U(\vec{x}) \sim 1/|\vec{x}|^{1+\epsilon}$. The important case of a Coulomb interaction corresponds to $\epsilon = 0$, while for a short-range interaction, $\epsilon = 1$, and the term (\ref{eq:Sphi}) is local. In case of the nematic QCP or spinon FS phase, the physical value of $\epsilon$ is $\epsilon = 1$. However, one may be able to access $\epsilon = 1$  via an expansion around $\epsilon = 0$.\cite{nayak, Mross} We, thus, work in the regime $0 \leq \epsilon \ll 1$ below.   Proceeding finally to the case of the  Mott transition, the screening of the gauge field by the gapless boson $b$ also generates a non-local term  (\ref{eq:bosonRPA}) in the gauge action, {\it i.e.} the effective action for the gauge-spinon sector is described by Eqs.~(\ref{eq:Lf})-(\ref{eq:Sphi}) with $\epsilon = 0$. 

In the case of the Ising nematic transition, the Lagrangian also allows for a perturbation $r \phi^2$, which tunes the system across the QCP. Below, we will work directly at the QCP, setting $r = 0$. We also perform all our RG calculations at temperature $T = 0$. As usual, we treat finite $T$ as an infra-red cut-off when running the RG equations.

As already noted, distinct pairs of patches $j \neq j'$ are decoupled in the above description and can be treated independently. We will shortly discuss the crucial role played by the inter-patch interactions in the pairing physics, however, for now, let us ignore such couplings and review the RG analysis of the two-patch theory (\ref{eq:Lf}) - (\ref{eq:Sphi}).\cite{nayak, Mross} The two-patch theory is described by a single dimensionless coupling constant, 
\beq
\alpha \equiv \frac{g^2 v_F \Lambda^{-\epsilon}_y}{(2\pi)^2}. 
\eeq
The fermion part of the action (\ref{eq:Lf}) dictates the scaling of frequency and momenta: 
\beq
\omega \to e^{-z_f \ell}  \omega, \quad q_\mx \to e^{-\ell} q_\mx, \quad q_\my \to e^{-\ell/2} q_{\my}, 
\eeq
with the bare dynamical exponent, $z_f = 1$. As we will see below, $z_f$ will generally be renormalized by interactions, however, the ``anisotropic" momentum scaling, $q_\mx \sim q_\my^2$, is exact due to the non-renormalization of the FS curvature $K$.\cite{MS} The full interacting fermion Green's function $G(\omega, \vec{q})$ depends only on the distance to the FS, $q_x + q^2_y/(2 K)$, so we may identify $z_f$ with the fermion dynamical exponent. On the other hand, the full boson propagator $D(\omega, \vec{q})$ of the two-patch theory depends only on the momentum tangent to the FS, $q_y$, so the above scaling fixes the relationship\cite{MS} between the boson dynamical exponent $z_b$ and the fermion dynamical exponent $z_f$,
\beq
z_b = 2 z_f.
\eeq

Under the above scaling with bare $z_f = 1$, $\alpha$ flows as ${d \alpha}/{d \ell} = ({\epsilon}/{2}) \alpha$. Hence, the fermion-boson interaction is irrelevant for $\epsilon < 0$, relevant for $\epsilon > 0$ and marginal at tree-level for $\epsilon = 0$.  To compute quantum corrections to the RG flow one can  utilize either a perturbative expansion in $\alpha$ (Ref.~\onlinecite{nayak}; however, see footnote \onlinecite{Loop}) or a $1/N$ expansion (Ref.~\onlinecite{Mross}).  At leading order both expansions give the same result. To one loop order (first order in $1/N$), $\alpha$ and $v_F$ run as,
\bea \frac{d \alpha}{d \ell} &=&  \frac{\epsilon}{2} \alpha - \frac{\alpha^2}{N}  \label{eq:RGalpha}\\
\frac{d v_F}{d \ell} &=& - \frac{\alpha}{N} v_F \label{eq:RGvF}\eea
and the fermion field acquires an anomalous dimension,
\beq f(\omega, q_\mx, q_\my) \to\left[1+ \left(\frac{7}{4} - \frac{\eta_f}{2}\right)d\ell\right] f(e^{d\ell}\omega, e^{d\ell} q_\mx, e^{d \ell/2} q_\my)  \eeq
with $\eta_f = {\alpha}/{N}$. For $\epsilon = 0$, $\alpha$ flows logarithmically to zero, and the system is a marginal Fermi-liquid with the fermion self-energy,
\beq  \Sigma(\omega) \sim - i \frac{\alpha}{N} \omega \log \frac{\Lambda_\omega}{|\omega|}\eeq 
with $\Lambda_\omega \sim v_F \Lambda_\mx$ - the energy cut-off.  For $\epsilon > 0$, the flow (\ref{eq:RGalpha}) has an infra-red stable fixed point at $\alpha_* = N \epsilon/2$. If $N$ is of $O(1)$ then $\epsilon \ll 1$ ensures that the fixed-point occurs at weak coupling. On the other hand, if $N \gg 1$, we take $\epsilon \sim O(1/N)$ to make $\alpha_* \sim O(1)$ and obtain a well-defined large-$N$ limit. In either case, at the fixed point, 
\beq \frac{d v_F}{d \ell} = - \frac{\epsilon}{2} v_F, \quad \eta_f = \frac{\epsilon}{2}  \label{eq:Floweps}\eeq
implying a fermionic dynamical exponent
\beq
z_f = 1+ \frac{\epsilon}{2}, \label{eq:zfeps}
\eeq
and a fermion self-energy 
\beq
\Sigma_f (\omega) \sim \omega^{1 -\eta_f}. 
\eeq
The exponent $z_f$ directly manifests itself in the nFL specific heat, \beq
C \sim T^{1/z_f}.
\eeq

We note that the expression for $z_f$ in Eq.~(\ref{eq:zfeps}) holds to all orders in $\epsilon$: this is tied to the non-analytic nature of the $q_\my$
dependence in $S_\phi$, which undergoes no renormalization. On the other hand, for $\epsilon = 1$, $S_\phi$ is analytic in $q_\my$ and, 
in principle, can undergo renormalization. Our ability to access the physically important $\epsilon = 1$ point through an expansion around $\epsilon = 0$ is, thus, tied to such renormalizations being
absent. No renormalization of $S_{\phi}$ in the $\epsilon = 1$ theory has been found up to three loop order, \cite{MS} however, a general proof of this statement is currently lacking.

\begin{figure*}[t]
\begin{center}
\includegraphics[width=6in]{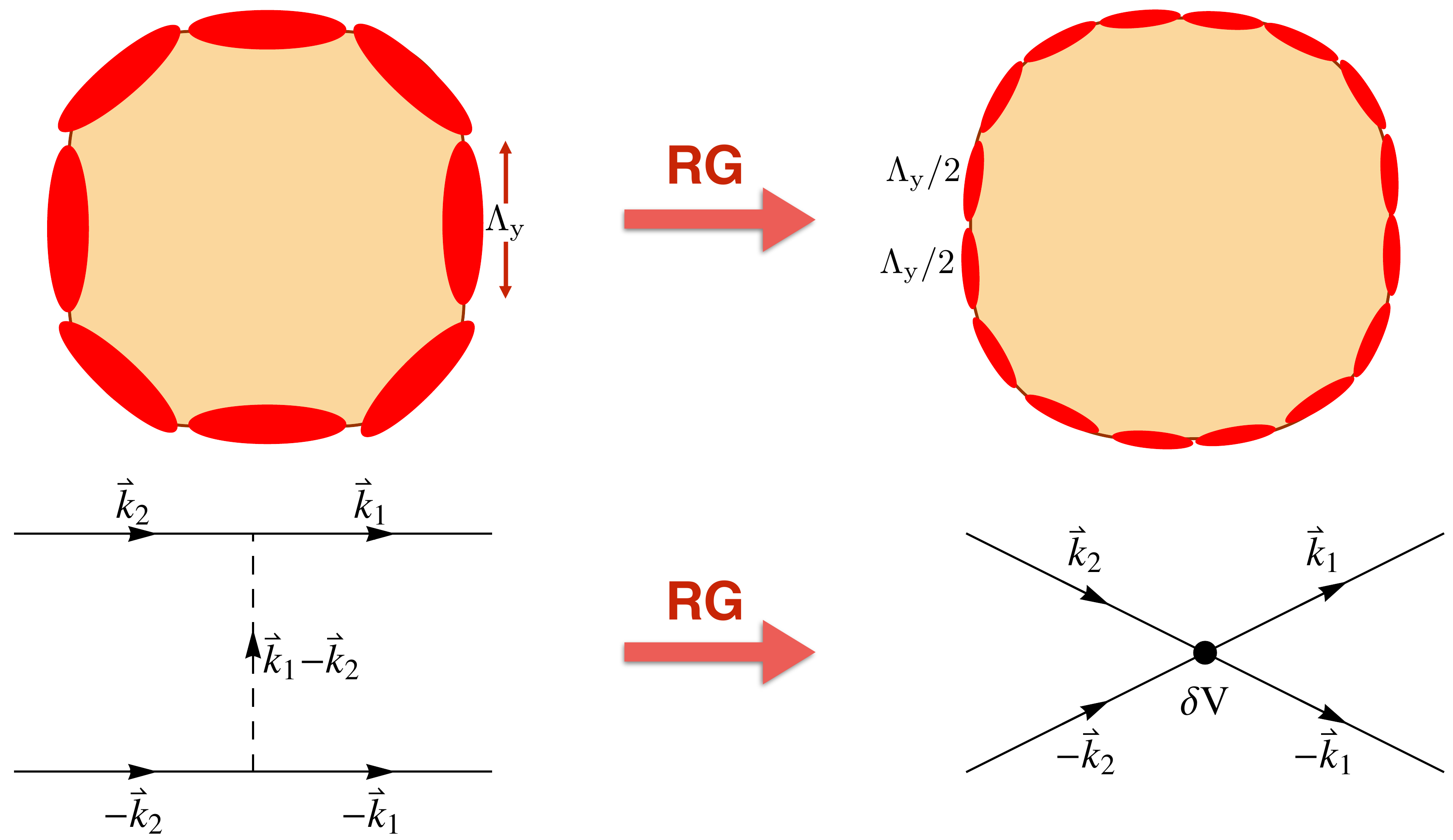}
\caption{ Top: Our RG procedure. During each RG step, each patch of the Fermi-surface is divided into two smaller patches. The relationship between the widths and heights of the patches remains as in Eq.~(\ref{lambdaxy}). Bottom: Single boson exchange mediates a non-local intra-patch interaction (left). Here and below, solid/dashed lines are fermion/boson propagators. As high momentum boson modes are integrated out in the RG process, a local inter-patch four-Fermi interaction in the BCS channel $\delta V(\vec{k}_1, -\vec{k}_1; \vec{k}_2, -\vec{k}_2)$ is generated (right).}
\label{fig:Collapse}
\end{center}
\end{figure*}
We next return to consider the effect of inter-patch interactions, which have been mostly ignored in previous studies. However, as we demonstrate below, such couplings must be included in the theory, as they are automatically generated in the RG process. This fact was first noted in Ref.~\onlinecite{sonqcd} in the context of 3d QCD at finite quark density, and here we closely follow the RG treatment proposed by Ref.~\onlinecite{sonqcd}. So far, we have left the precise RG procedure somewhat implicit. Recall that under the scaling we advocated for the two-patch theory, $q_\mx \to e^{-\ell} q_\mx$ and $q_\my \to e^{-\ell/2} q_\my$, so in the RG process we reduce both the fermion momentum cut-off perpendicular to the FS, $\Lambda_\mx$, and the cut-off tangent to the FS, $\Lambda_\my$. While $\Lambda_\mx$ can be, as usual, shrunk by integrating out gapped fermion excitations away from the FS, reducing $\Lambda_\my$ in the same manner would require integration over gapless fermions {\it on} the FS, which is illegal. Instead, during each RG step we re-partition the FS into smaller patches with width $\Lambda'_\my = e^{-\ell/2} \Lambda_\my$, while the reduction in the patch thickness $\Lambda'_\mx = e^{-\ell} \Lambda_\mx$  is still performed by integrating out gapped fermions away from the FS. Simultaneously, in each RG step we integrate out boson fluctuations with momenta $e^{-\ell/2} \Lambda_\my < |\vec{q}| < \Lambda_\my$;
see Fig.~\ref{fig:Collapse}. Before the RG step, such boson fluctuations mediate non-local {\it intra}-patch interactions between the fermions. However, after the RG step, these generate a local four-fermion {\it inter}-patch interaction, as shown in Fig.~\ref{fig:Collapse} (bottom).

As is well known from ordinary FL theory, a very restricted set of four-fermion {\it inter}-patch couplings on the FS is kinematically allowed.\cite{Shankar, PolchinskiTASI} Only forward-scattering and BCS scattering interactions survive as the shell of fermion states around the FS is shrunk in the RG process. As we are interested in the physics of pairing, in the present paper we concentrate only on four-fermion interactions in the BCS channel, which can be described by the action,
\begin{widetext}
\bea S^{\mathrm{BCS}} &=& -\frac{1}{4} \int \prod_{i =1}^{4} \frac{d^2 \vec{k}_i d \omega_i }{(2\pi)^3} f^{\dagger}_\alpha(k_1) f^{\dagger}_\beta(k_2) f_\gamma(k_3) f_\delta(k_4) 
(2 \pi)^3 \delta^3(k_1 + k_2 - k_3 - k_4) \nn\\
&& ~~~\times \left((\delta_{\alpha \gamma} \delta_{\beta \delta} + \delta_{\alpha \delta} \delta_{\beta \gamma}) V^a(\vec{k}_1, \vec{k}_2; \vec{k}_3, \vec{k}_4) + (\delta_{\alpha \gamma} \delta_{\beta \delta} - \delta_{\alpha \delta} \delta_{\beta \gamma}) V^s(\vec{k}_1, \vec{k}_2; \vec{k}_3, \vec{k}_4)\right)\label{eq:BCS} \eea
\end{widetext}
Here, $V^s$/$V^a$ are, respectively, symmetric/antisymmetric under exchanging $\vec{k}_1 \leftrightarrow \vec{k}_2$, and $\vec{k}_3 \leftrightarrow \vec{k}_4$. Only the values of the interaction for BCS-matched momenta, $V^{s,a}(\vec{k}_1, -\vec{k}_1; \vec{k}_2, -\vec{k}_2)$, play a role; furthermore, $\vec{k}_{1,2}$ can be taken to lie on the FS. From now on, we assume that the system is rotationally invariant,\cite{FNcircular} so we may write $V^{s,a}(\vec{k}_1, - \vec{k}_1; \vec{k}_2, - \vec{k}_2) = V^{s,a}(\theta_1 - \theta_2)$, with $\theta_{1,2}$ - angles on the FS. Performing an expansion in angular harmonics,
\beq V^{s,a}(\theta_1-\theta_2) = \sum_{m = -\infty}^{\infty} V^{s,a}_m e^{i m (\theta_1 - \theta_2)}\eeq
$V^s$ involves only even angular momentum components and $V^a$ - odd. It is convenient to define dimensionless BCS interaction constants, \beq
\widetilde{V}^{s,a}_m \equiv \frac{k_F}{2 \pi v_F} V^{s,a}_m. \label{deftildeV}
\eeq

\begin{figure}[t]
\begin{center}
\includegraphics[width=3.5in]{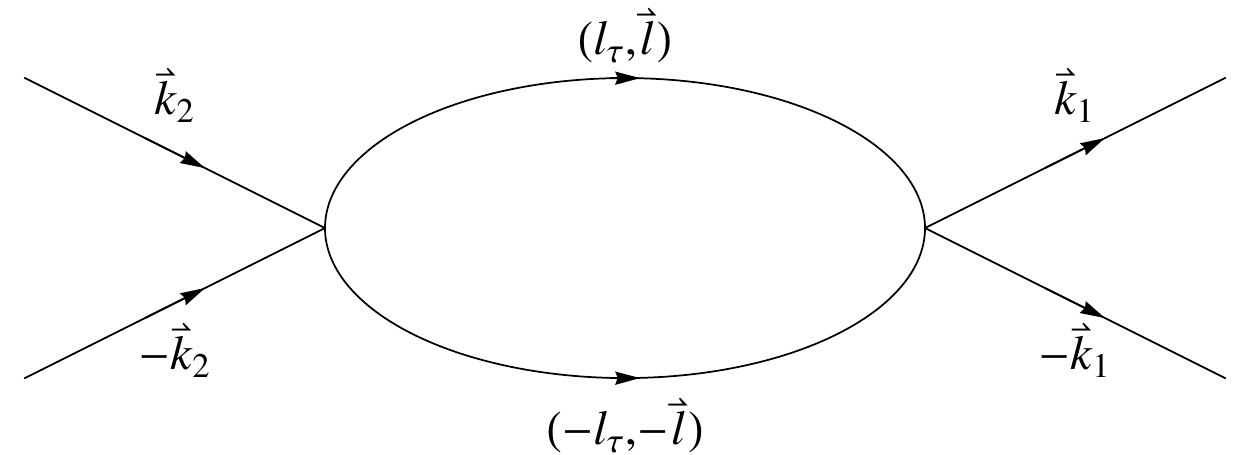}
\caption{Renormalization of the BCS interaction $V(\vec{k}_1, -\vec{k}_1;\vec{k}_2, - \vec{k}_2)$ in a FL.}
\label{fig:BCS}
\end{center}
\end{figure}

In the absence of the coupling to the gapless boson ({\em i.e.\/} in a Fermi-liquid), the RG flow of the 
BCS interaction (\ref{eq:BCS}) can be determined as in Refs.~\onlinecite{Shankar, PolchinskiTASI}.
The RG in their work involves only the rescaling of $\Lambda_\mx$, which is the same as that in Fig.~\ref{fig:Collapse} (top).
Our rescaling of $\Lambda_\my$ plays no role in the renormalization of the BCS interaction, and so we can read off
the renormalization of $\widetilde{V}^{s,a}_m$ from their results: this interaction
is marginal at tree level, and acquires the following flow at one-loop level (see Fig.~\ref{fig:BCS}), 
\beq \frac{d \widetilde{V}^{s,a}_m}{d \ell} = -  (\widetilde{V}^{s,a}_m)^2 \label{eq:RGFL}\eeq
Thus, in a Fermi-liquid, if the initial value of the BCS interaction is repulsive, $V^{s,a}_m > 0$, then $V^{s,a}_m$ flows logarithmically to zero, while if the initial value of the interaction is attractive, $V^{s,a}_m < 0$, $V^{s,a}_m$ runs away to $-\infty$ at an energy scale, $\Delta_{\mathrm{BCS}} \sim \Lambda_\omega \exp(-1/|\widetilde{V}^{s,a}_m|)$, signaling an instability to fermion pairing.

Next, we study how the flow of the four-fermion BCS interactions (\ref{eq:RGFL}) is modified by the presence of the gapless boson $\phi$. In the limit $\alpha \ll 1$ (or $N \gg 1$),  $\widetilde{V} \ll 1$, the leading modification comes from the diagram in Fig.~\ref{fig:Collapse} (bottom, left), which represents the one-boson exchange contribution to the four-fermion BCS amplitude.  As already noted, integration over intermediate large-momentum $\phi$ modes in Fig.~\ref{fig:Collapse}  generates an {\it inter}-patch four-fermion interaction,
\beq \delta V^{s,a}(\vec{k}_1, -\vec{k}_1; \vec{k}_2, -\vec{k}_2) = - \frac{\zeta}{2} v^2_F D_>(0, \vec{k}_1 - \vec{k}_2) \label{eq:VD}\eeq
where $D(\omega, \vec{q})$ is the boson propagator and the subscript ``$>$" indicates that only modes in the momentum shell $e^{-\ell/2} \Lambda_{\my} < |\vec{q}| < \Lambda_{\my}$ should be kept.  
We remind the reader that the constant $\zeta$ distinguishes between the different nFLs: we have $\zeta=1$ for the Ising-nematic case,
and $\zeta=-1$ for the spinon Fermi-surface phase/Mott transition and HLR cases.
Note that the frequencies of the external fermions and, hence, of the boson in Fig.~\ref{fig:Collapse} (bottom) can be set to 0. Eq.~(\ref{eq:VD}) gives $\delta V$ for the case of small angle scattering, $\vec{k}_1 \to \vec{k}_2$; the result for $\vec{k}_1 \to - \vec{k}_2$ is determined by symmetry.  The static boson propagator is given by $D(0, \vec{q}) = {g^2}/({N |\vec{q}|^{1+\epsilon}})$. Computing the angular harmonics corresponding to (\ref{eq:VD}),
\bea \delta \widetilde{V}^{s,a}_m &=& - 2 \left(\frac{k_F}{2 \pi v_F}\right) \frac{\zeta}{2} v^2_F  \int \frac{d \theta}{2\pi} D_>(0, k_F \theta) e^{- i m \theta} \nn \\
&=& -\frac{\zeta g^2 v_F}{2 \pi^2 N} \int_{e^{-\ell/2} \Lambda_y}^{\Lambda_y} \frac{d q}{q^{1+\epsilon}} \cos(m q/k_F) = - \zeta \frac{\alpha}{N} \ell\nn\\\eea
In the last step, we have dropped the factor $\cos(m q/k_F)$ as $\Lambda_y \ll k_F$. Thus, the process in Fig.~\ref{fig:Collapse} (bottom) contributes a term ${d \widetilde{V}^{s,a}_m}/{d \ell} = -\zeta \alpha/N$ to the RG flow of $V$, which combines with Eq.~(\ref{eq:RGFL}) to give,
\beq \frac{d \widetilde{V}^{s,a}_m}{d \ell} = -\zeta \frac{\alpha}{N} -  (\widetilde{V}^{s,a}_m)^2 \label{eq:FlowV}\eeq
There are also terms of order $\alpha \widetilde{V}^{s,a}_m$ which arise from vertex corrections and the flow of $v_F$ in the definition
(\ref{deftildeV}), but we have dropped them because they are are higher order in $\epsilon$. 
Note that the flow (\ref{eq:FlowV}) is independent of the angular momentum and spin channel; hence we drop the angular momentum/spin indices on $V$ below. The flow equation (\ref{eq:FlowV}) for the {\it inter}-patch BCS interaction in conjunction with the flow of the {\it intra}-patch coupling constant $\alpha$ in Eq.~(\ref{eq:RGalpha}) determines the physics of the nFL states considered. We next analyze these RG equations and discuss their consequences. However, we first point out that in the regime of analytical control $\epsilon \ll 1$, all the conclusions of our RG treatment can be reproduced by solving the Eliashberg equation for the pairing vertex, as has been
shown elsewhere.\cite{MMpaper} This lends further support to our results.

\section{Results: Ising-nematic QCP} 
\label{sec:NematicRes}
We first discuss the solution to RG Eqs.~(\ref{eq:RGalpha}), (\ref{eq:FlowV}) for the nematic QCP. In this case, the constant $\zeta = 1$ in Eq.~(\ref{eq:FlowV}), so the fluctuations of the order parameter captured by the first term in Eq.~(\ref{eq:FlowV}) drive the short-range interaction $V$ negative (attractive), as expected. In fact, as discussed in appendix \ref{sec:RGnem}, we find that independent of the initial values of $\alpha$ and $V$, $V$ flows  to $-\infty$ at a finite $\ell = \ell_p$, indicating an instability of the Ising-nematic QCP to superconductivity.  Thus, we expect both the zero temperature electron pairing gap $\Delta$ and the superconducting critical temperature $T_c$ to be proportional to $\Delta \sim T_c \sim \Lambda_\omega e^{-\ell_p}$.  Unlike in the ordinary Fermi-liquid, the run-away flow $V \to -\infty$ occurs even if the initial value of $V$ is repulsive: gapless order parameter fluctuations eventually drive $V$ attractive. However, the magnitude of $\ell_p$ and hence the pairing gap does depend on the initial value of $V$: the smaller the initial $V$ - the larger the gap.  As already noted, the flow equations for $V^{s,a}_m$  in different angular momentum/spin channels decouple and are identical. We, thus, expect pairing to occur in the channel where $V^{s,a}_m$ diverges first, {\em i.e.\/} one which has the smallest initial $V^{s,a}_m$. Hence, the pairing symmetry is non-universal.

It is interesting to compare the pairing scale $\Delta$ with the energy scale $E_{\mathrm{nFL}} = \Lambda_\omega e^{-\ell_{\mathrm{nFL}}}$ at which electronic quasiparticles get destroyed. Here, we identify $E_{\mathrm{nFL}}$ as the energy at which the Fermi-velocity $v_F$, whose flow is determined by Eq.~(\ref{eq:RGvF}), starts to deviate significantly from its bare value. We find that as long as our calculation is controlled, ({\em i.e.\/} $\epsilon \ll 1$), $E_{\mathrm{nFL}} \ll \Delta$, so the superconducting instability preempts the destruction of quasiparticles and associated nFL behavior. This is quite distinct from the physics of many materials where nFL behavior is observed at energies/temperatures well above the superconducting $T_c$. As we take the artificial control parameter $\epsilon$ to its physical value $\epsilon = 1$, the two scales $E_{\mathrm{nFL}}$ and $\Delta$ approach each other, however, at this point we lose analytical control.

We now briefly illustrate the above conclusions for several regimes of $\epsilon$, $\alpha$, $V$ (see appendix \ref{sec:RGnem} for more details).  First, consider the case $\epsilon = 0$. Here, we find
\beq \Delta = \Lambda_\omega \exp\left[-\frac{1}{\sqrt{\widetilde{\alpha}}} \left(\frac{\pi}{2} + \tan^{-1}\frac{\widetilde{V}}{\sqrt{\widetilde{\alpha}}}\right)\right] \label{eq:ell1}\eeq
with 
\beq
\widetilde{\alpha} \equiv \frac{\alpha}{N}. 
\eeq
If the bare short-range interaction $\widetilde{V}$ is small compared to the long-range interaction, $|\widetilde{V}| \ll \sqrt{\widetilde{\alpha}}$, then Eq.~(\ref{eq:ell1}) reduces to $\Delta = \Lambda_\omega \exp(-\pi/(2 \sqrt{\widetilde{\alpha}}))$. On the other hand, if the bare short-range interaction $\widetilde{V}$ is large and repulsive, $\widetilde{V} \gg \sqrt{\widetilde{\alpha}}$, $\Delta = \Lambda_\omega \exp(-\pi/ \sqrt{\widetilde{\alpha}})$, {\em i.e.\/} the gap is reduced by a factor of two on the logarithmic scale compared to the case of small $\widetilde{V}$. Finally, if the bare short-range interaction is large and attractive, $\widetilde{V} < 0, |\widetilde{V}| \gg \sqrt{\widetilde{\alpha}}$, the gap takes the standard BCS form, $\Delta = \Lambda_\omega \exp(-1/|\widetilde{V}|)$. The scale at which nFL effects become appreciable is $E_{\mathrm{nFL}} \sim \Lambda_\omega \exp(-1/\widetilde{\alpha})$. Thus, as long as $\widetilde{\alpha} \ll 1$, the pairing gap $\Delta$ is parametrically larger than the nFL scale $E_{\mathrm{nFL}}$.  We note in passing that the result (\ref{eq:ell1}) is identical to one obtained for the problem of quark pairing by color gauge fields in 3d dense baryonic matter,\cite{sonqcd} and for electron pairing near a ferromagnetic QCP in 3d.\cite{ChubukovSchmalian}

Proceeding to the case $\epsilon > 0$ (which may be continuously connected to the physical case $\epsilon = 1$), we find that the nFL scale is still given by $E_{\mathrm{nFL}} \sim \Lambda_\omega \exp(-1/\widetilde{\alpha})$ for $\widetilde{\alpha} \gg \epsilon$, as well as for $\widetilde{\alpha} \sim O(\epsilon)$, while for $\widetilde{\alpha} \ll \epsilon$, $E_{\mathrm{nFL}} \sim \Lambda_\omega (\tilde{\alpha}/\epsilon)^{2/\epsilon}$. The pairing scale $\Delta$ is still given by the expression in Eq.~(\ref{eq:ell1}) for $\widetilde{\alpha} \gg \epsilon^2$, so the relation $E_{\mathrm{nFL}} \ll \Delta$ holds. For $\widetilde{\alpha} \ll \epsilon^2$ and $\widetilde{V} > 0$ (or $\widetilde{V} < 0$, but $|\widetilde{V}| \ll \epsilon/\log \frac{\epsilon^2}{\widetilde{\alpha}}$), we obtain 
\beq \Delta \sim \Lambda_\omega \left(\frac{\widetilde{\alpha}}{\epsilon^2}\right)^{2/\epsilon} \label{eq:Deltapower}\eeq
so $\Delta$ depends on the coupling constant $\tilde{\alpha}$ in a power-law manner and $E_{\mathrm{nFL}}/\Delta \sim \epsilon^{2/\epsilon} \ll 1$. Eq.~(\ref{eq:Deltapower}) has been previously obtained within an Eliashberg-like treatment in Ref.~\onlinecite{Yuzbashyan}. Naive extrapolation of the above result to the physically relevant value $\epsilon = 1$ gives, $\Delta \sim E_{\mathrm{nFL}} \sim \tilde{\alpha}^2 \Lambda_\omega$, {\em i.e.\/} the pairing and nFL scales become parametrically equal. This conclusion is again supported by the direct solution of Eliashberg-like equations.\cite{BonesteelNayak,ChubukovMoon}  

\begin{figure}[t]
\begin{center}
\includegraphics[width=3.5in]{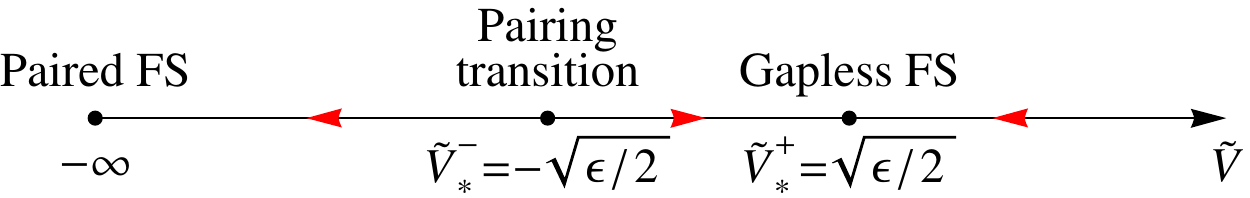}
\caption{RG flow of the inter-patch BCS interaction $\widetilde{V}^{s,a}_m$ in the spinon FS/HLR phase with $0 < \epsilon \ll 1$. The IR stable fixed-point $\widetilde{V}^+_* \approx \sqrt{\epsilon/2}$ controls the gapless FS phase, while the IR unstable fixed point $\widetilde{V}^-_* \approx -\sqrt{\epsilon/2}$ controls the continuous transition to the paired phase.}
\label{fig:Vflow1}
\end{center}
\end{figure}

Note that in the above analysis, we have ignored the $d$-wave dependence of the coupling between the Ising-nematic order parameter and electrons on the angle around the FS. We don't expect the $d$-wave form-factor to affect the maximum magnitude of the pairing gap strongly, however, it will certainly affect its angular dependence. In fact, recent results of Maier and Scalapino\cite{ScalapinoNem} and Lederer {\it et al.}\cite{KivelsonNem} suggest that the angular dependence of the gap at the QCP may be quite singular. These authors study the regime where the system is tuned sufficiently away from the Ising-nematic QCP that the standard weak coupling BCS machinery can be applied. 
They find that as the QCP is approached, the superconducting gap becomes strongly peaked around the angle where the coupling between the order parameter and the electrons is maximal ({\em i.e.} around the anti-node). It is interesting whether this result survives all the way to the QCP. In the future, we hope to settle this question by extending our RG analysis to the physical case with no rotational symmetry.

While our RG analysis is performed exactly at the metallic critical point, superconductivity will survive when one tunes the system slightly away from the QCP with the perturbation $r \phi^2$. Recall that $r$ induces a finite correlation length for the order parameter, $\xi_{\phi} \propto r^{-\nu}$, $\nu = (1+\epsilon)^{-1}$, with the corresponding energy scale $E_{\phi}  \sim \xi^{-z_b}_\phi \sim r^{(2+\epsilon)/(1+\epsilon)}$. Away from the QCP, $E_\phi$ serves as an IR cut-off on the RG equation (\ref{eq:FlowV}). Therefore, the pairing gap $\Delta(r)$ will be essentially unmodified from its QCP value $\Delta_0 = \Delta(r = 0)$, as long as $E_\phi(r) \ll \Delta_0$. We may then estimate the characteristic width of the superconducting dome as $\delta r \sim {\Delta_0}^{(1+\epsilon)/(2+\epsilon)}$ (here, we consider the most interesting regime when pairing at the QCP is dominated by order parameter fluctuations rather than the bare short-range BCS attraction $V$). As always, the precise shape of the dome for $|r| \sim \delta  r$ cannot be determined from RG considerations alone. The dome will generally have tails extending to $r \gg \delta r$, where the gap ($T_c$) will be strongly suppressed compared to $\Delta_0$. The precise form of $\Delta(r)$ in these tails can be obtained by running the RG equation (\ref{eq:FlowV}) for the BCS coupling up to the energy scale $E_\phi$. It is easy to see that for $r \gg \delta r$, $|\tilde{V}(E_\phi)| \ll 1$, {\it i.e.} the system at energy $E_\phi$ is in the weak-coupling regime.  Below the energy $E_\phi$, $\phi$ is not critical, and the system is described by Fermi-liquid theory, so the BCS coupling continues to flow according to Eq.~(\ref{eq:RGFL}). Therefore, if $\tilde{V}(E_\phi) < 0$, then the system will develop superconductivity, with $\Delta(r) = E_\phi \exp(-1/|\tilde{V}(E_\phi)|)$. On the other hand, if $\tilde{V}(E_\phi) > 0$, no superconducting instability will occur. Thus, if the bare $V$ is attractive, the tails of the superconducting dome will extend to all $r$ (as long as one remains in the regime of applicability of the critical theory). On the other hand, if the bare $V$ is repulsive, then the dome will terminate at a finite $r$, corresponding to $V(E_\phi) = 0$. 

We would like to note that in the above discussion, $r$ denoted the deviation from the ``metallic" QCP. As we saw, this QCP is unstable to superconductivity, so the true Ising-nematic QCP will occur inside the superconducting dome. In addition, its location will generally shift away from that of the putative metallic QCP at $r = 0$ to $r = r_c$. On general scaling grounds, we expect $|r_c| \sim \delta r$. The universality class of the true QCP at $r  = r_c$ depends on whether pairing gives rise to a fully gapped or a nodal superconductor. If the superconductor is fully gapped, this transition will be in the classical 3D Ising univerality class. The character of this transition in a nodal $d$-wave superconductor has been discussed in Ref.~\onlinecite{YejinSubir}. The critical behavior associated with the true QCP will only be observable for $T \ll \Delta_0$, and the system will cross over to the metallic critical behavior discussed in this paper for $T \gg \Delta_0$.

\section{Results: Spinon FS phase and HLR phase} 
\label{sec:resSFS}
We now turn to the solution of the RG equations Eqs.~(\ref{eq:RGalpha}), (\ref{eq:FlowV}) for the spinon FS phase and HLR phase. The constant $\zeta$ in Eq.~(\ref{eq:FlowV}) now takes the value $\zeta = -1$, hence gauge field fluctuations drive $\widetilde{V}$ repulsive, in accordance with intuition. We first solve Eqs.~(\ref{eq:RGalpha}), (\ref{eq:FlowV}) when $\epsilon > 0$ (with an eye to describing the physical spinon FS phase and the HLR phase with short-range interactions, where $\epsilon = 1$). Here, the coupling $\alpha$ flows to the fixed point $\alpha_* = N\epsilon/2$, and we may substitute this fixed point value into the RG equation for $V$, (\ref{eq:FlowV}). We then find two perturbatively accessible fixed points for $\widetilde{V}$: $\widetilde{V}^{\pm}_* = \pm \sqrt{\epsilon/2}$, see Fig.~\ref{fig:Vflow1}. The fixed point $\widetilde{V}^+_*$ is infra-red stable; as long as the initial value of $\widetilde{V}$ is greater than $\widetilde{V}^-_*$, $\widetilde{V}$ flows to $\widetilde{V}^+_*$. Thus, the spinon FS and HLR phases are controlled by the fixed point $(\alpha_*, \widetilde{V}^+_*)$ and are ${\it stable}$ to fermion pairing. However, if the initial value of $\widetilde{V}^{s,a}_m$ in some angular momentum/spin channel is smaller than $\widetilde{V}^-_*$, $\widetilde{V}^{s,a}_m$ runs away to $-\infty$, and fermion pairing occurs. $\widetilde{V} = \widetilde{V}^-_*$, thus, marks the phase transition between the $U(1)$ and $\mathbb{Z}_2$ spin-liquid phases (HLR and incompressible QH phases). Note that unlike in a Fermi-liquid, a finite strength of the attractive short-range interaction $|\widetilde{V}| > |\widetilde{V}^-_*| > 0$ is needed to overcome the long-range repulsion mediated by the gauge field and cause fermion pairing. Pairing in a given angular-momentum/spin channel can be driven by tuning the corresponding $\widetilde{V}^{s,a}_m$. The pairing transition is continuous and the spinon (neutral fermion) gap onsets in a power law fashion, $\Delta \sim (\widetilde{V}^-_* - \widetilde{V})^{z \nu}$, where
\beq  \frac{1}{z \nu} = \frac{d}{d \widetilde{V}}\left( \frac{d\widetilde{V}}{d \ell}\right)\bigg|_{\widetilde{V}  = \widetilde{V}^-_*} = \sqrt{2 \epsilon}\eeq
This is, again, distinct from an ordinary FL where the electron gap has the familiar exponential form $\Delta \sim \exp(-1/|\widetilde{V}|)$. 

We note that to the leading order in $\epsilon$ discussed above, the presence of inter-patch interactions $V$ does not affect the flow of the intra-patch coupling constant $\alpha$, Eq.~(\ref{eq:RGalpha}), and the Fermi-velocity $v_F$, Eq.~(\ref{eq:RGvF}).   As a result, most physical properties (fermion and boson dynamical exponents $z_f$, $z_b$; specific heat; $2 k_F$ exponents\cite{Mross} etc.), at the two fixed points $V = V^{\pm}_*$ are identical. This conclusion may be true to all orders in $\epsilon$, since, perturbatively, BCS interactions do not influence the single particle properties ($v_F$, $Z$) in a FL.

We next discuss the marginal case $\epsilon = 0$, which describes the QH fluid with Coulomb interactions. Here, the coupling constant $\alpha$ logarithmically flows to $0$. The combined flow of $\widetilde{\alpha}$, $\widetilde{V}$ is shown in Fig.~\ref{fig:Vflow2} (see appendix \ref{sec:RGHLR} for details). The flow is characterized by a single fixed-point $\widetilde{\alpha} = 0$, $\widetilde{V} = 0$ and features an attractor line $\widetilde{V} = \sqrt{\widetilde{\alpha}}$ and a separatrix $\widetilde{V} = -\sqrt{\widetilde{\alpha}}$. As long as the initial values of $\widetilde{V}$, $\widetilde{\alpha}$ satisfy $\widetilde{V} > -\sqrt{\widetilde{\alpha}}$, the couplings flow to the attractor line $\widetilde{V}= \sqrt{\widetilde{\alpha}}$ and then into the fixed point $\widetilde{\alpha} = 0$, $\widetilde{V} = 0$. So, the HLR phase with Coulomb interactions is stable in a finite region of parameter space. On the other hand, if the initial $\widetilde{V} < -\sqrt{\widetilde{\alpha}}$, $\widetilde{V}$ runs away to $-\infty$ and fermion pairing occurs. Thus, the separatrix $\widetilde{V} = -\sqrt{\widetilde{\alpha}}$ describes the transition between the HLR phase and the paired QH phase. Note that this separatrix also logarithmically flows into the fixed point $\widetilde{\alpha} = 0$, $\widetilde{V} = 0$, so the stable and the unstable fixed points $\widetilde{V}^{\pm}_*$, found for $\epsilon > 0$, merge into a single fixed-point here. The pairing transition is continuous and the fermion gap turns on as the separatrix is crossed in an unusual super-power law fashion,
\beq \Delta \sim \exp\left[-\frac{1}{16} \log^2 (V_c- V)\right]\eeq
with $V_c \approx -\sqrt{\widetilde{\alpha}}$.

\section{Results: Mott transition}
\begin{figure}[t]
\begin{flushleft}
\includegraphics[width=4in,trim =90 0 0 0,clip]{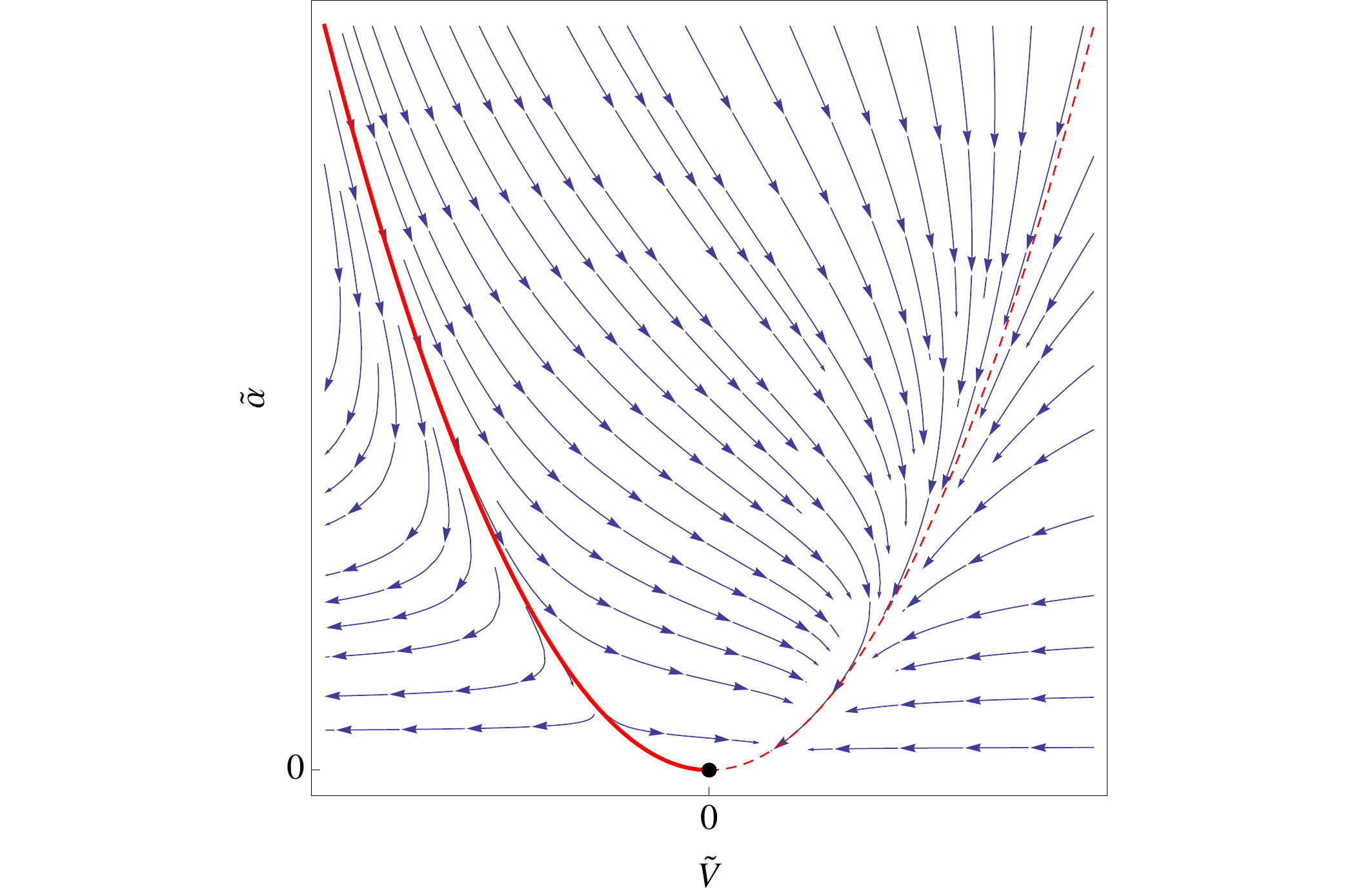}
\caption{RG flow of the intra-patch coupling constant $\widetilde{\alpha}$ and the inter-patch BCS interaction $\widetilde{V}^{s,a}_m$ in the HLR phase with Coulomb interactions ($\epsilon = 0$). Note the attractor line $\widetilde{V} \approx \sqrt{\widetilde{\alpha}}$  (dashed red curve) and the separatrix $\widetilde{V} \approx -\sqrt{\widetilde{\alpha}}$ (solid red curve). The HLR phase is controlled by the logarithmic flow of the attractor line into the fixed point $(\widetilde{V}, \widetilde{\alpha}) = (0,0)$. The phase transition to the paired CF phase is controlled by the logarithmic flow of the separatrix into the same fixed point $(\widetilde{V}, \widetilde{\alpha}) = (0,0)$.}
\label{fig:Vflow2}
\end{flushleft}
\end{figure}

As already noted, the $\epsilon = 0$ theory also describes the QCP between the spinon FS phase and a Fermi-liquid phase. Thus, the results in section \ref{sec:resSFS} imply that the spinon Fermi-surface at the Mott transition is stable as long as the initial values of $(\widetilde{V}, \widetilde{\alpha})$ lie to the right of the separatrix in Fig.~\ref{fig:Vflow2}. On the other hand, if the initial values of $(\widetilde{V}, \widetilde{\alpha})$ lie to the left of the separatrix, the spinon acquires a gap, and the Mott transition occurs between a $Z_2$ spin-liquid and a superconductor.

In the former regime $\widetilde{V} > -\widetilde{\alpha}$, where the spinon FS at the Mott transition is stable, we expect that an even stronger statement holds: the spinon Fermi-surface remains stable as one tunes the system slightly away from the Mott transition. Indeed, if one tunes the system into the compressible phase, $t < t_c$ in Eq.~(\ref{eq:Lb}), the gauge field becomes Higgsed by the condensate $\langle b \rangle \neq 0$ below a momentum scale $q_a \sim (t_c - t)^{\nu}$, where $\nu$ is the correlation length exponent of the XY universality class. The corresponding energy scale $E_a \sim q^2_a$ will serve as an IR cut-off on the RG equations for the flow of ($\tilde{V}$, $\tilde{\alpha}$), (\ref{eq:RGalpha}), (\ref{eq:FlowV}). Below this energy scale, gauge fluctuations become non-critical and the spinon FS will be described by FL theory. Now, as we discussed in section \ref{sec:resSFS}, for energies above $E_a$, the flow of $(\widetilde{V}, \widetilde{\alpha})$ tends to the attractor line $\tilde{V}(\ell) = \sqrt{\tilde{\alpha}(\ell)} \approx \ell^{-1/2} > 0$. Thus,  at the crossover scale $\tilde{V}(E_a) > 0$, so no spinon pairing will occur as one further lowers the energy into the Fermi-liquid regime. Consequently, the Fermi-liquid phase adjacent to the Mott transition will not develop superconductivity.

Likewise, if one tunes the system into the insulating phase, $t > t_c$ in Eq.~(\ref{eq:Lb}), the screening (\ref{eq:bosonRPA}) of the gauge field by $b$ will cease at a momentum scale $q_a \sim (t-t_c)^{\nu}$, with the corresponding energy scale $E_a \sim q^2_a$. Below this energy scale the system is effectively in the spinon FS phase. Again, by the time the scale $E_a$ is reached, $(\widetilde{V}, \widetilde{\alpha})$ will approach the attractor line $\tilde{V}(\ell) = \sqrt{\tilde{\alpha}(\ell)} \approx \ell^{-1/2} > 0$. Since, as we discussed in section \ref{sec:resSFS}, a finite strength of attraction $\tilde{V} < \tilde{V}^-_*$ is needed to destabilize the spinon Fermi-surface phase, we conclude that no spinon pairing will occur on the insulating side of the transition, as well.

 \vspace{0.5cm}

\section{ $\mathbb{Z}_2$ spin-liquid and QH states near the pairing transition} 
As we showed in section \ref{sec:resSFS}, spinon FS and HLR phases can be driven through a continuous pairing transition. We now comment on some properties of the paired phase in the vicinity of the transition.  In many ways, these paired states are analogous to ordinary superconductors. As we already noted, the paired phase supports two kinds of fundamental excitations: spinons/neutral fermions and vortices of the gauge field. The latter are visons of the $\mathbb{Z}_2$ spin-liquid/charge $e/4$ excitations of the paired CF phase. The vortex excitations are, thus, particularly important in the QH context as their energy determines the charge gap. So far, we have only determined the scaling of the fermion gap $\Delta$ near the pairing transition. We now crudely estimate the magnitude of the vortex gap. The fermion pair-condensate is suppressed in the vortex core, whose radius we take to be the  fermion correlation length $\xi \sim \Delta^{-1/z_f}$. Thus, the vortex gap $E_v \sim (\epsilon_n - \epsilon_p) \xi^2$, where $\epsilon_n - \epsilon_p$ is the energy density difference between the ``normal" phase and the paired phase. The scaling of the energy density at the pairing transition is $\epsilon \sim \omega^{1+1/z_f}$ (e.g. recall the specific heat $C \sim T^{1/z_f}$ both in the gapless FS phase and at the pairing transition), so setting the characteristic energy scale $\omega$ in the paired phase  to the fermion gap $\Delta$, $\epsilon_n - \epsilon_p \sim \Delta^{1 + 1/z_f}$ and $E_v \sim \Delta^{1-1/z_f}$. Therefore,  the vison/charge gap vanishes as one approaches the de-pairing transition, although more slowly than the spinon/neutral fermion gap.  For the physically interesting case of the spinon FS or the QH system with short-range electron-electron interactions, $\epsilon = 1$, $z_f = 3/2$ and $E_v \sim \Delta^{1/3}$. Note that our estimate of the vortex gap strictly only applies to the case $\epsilon > 0$, for $\epsilon = 0$, $z_f = 1^+$ and we expect $E_v$ to vanish logarithmically as $\Delta \to 0$.

As is well-known, superconductors can be classified as type-I or type-II depending on their response to an external (orbital) magnetic field $H$. Both types of superconductors are characterized by a Meissner effect (full expulsion of magnetic flux) for small $H$. As the magnetic field is increased, a (3d bulk) type-I superconductor undergoes a 1st order transition to a fully normal state at a critical value $H = H_c$. On the other hand, in a type-II superconductor, an array of Abrikosov vortices is induced for magnetic fields $H > H_{c1}$ and the normal state is recovered only for $H > H_{c2} > H_{c1}$. The type of a conventional superconductor is determined by the ratio of the electron correlation length $\xi$ and the magnetic penetration depth $\lambda$. For $\lambda \ll \xi$, the superconductor is type-I, while for $\lambda \gg \xi$ - it is type-II. 

Related ``typology" also exists in paired spinon/composite fermion phases.\cite{Sid1,Sid2}.  However we first need to understand what plays the role of the external magnetic field $H$ in these systems. In the quantum Hall case, the flux of the emergent magnetic field is simply the electron density. Thus, the analog of the external magnetic field is the electron chemical potential $\mu$. 
For the spinon FS phase on the triangular lattice, based on symmetry considerations, we expect an external orbital magnetic field $H$ to couple linearly to the flux of the emergent gauge field, $\nabla\times \vec{a}$: $\delta L = -\beta H (\nabla \times \vec{a})$, with $\beta$ - a coupling constant.\cite{MotrunichChir}.  Recall that the emergent gauge flux is physically identified with the spin-chirality $\vec{S}_1 \cdot (\vec{S}_2 \times \vec{S}_3)$ of the elementary triangles.\cite{MotrunichChir} Moreover, starting with the electron Hubbard model on the triangular lattice, in the insulating limit $t \ll U$, one finds that a coupling of the external orbital magnetic field to the spin-chirality is, indeed, induced at order $t^3/U^2$, so $\beta \sim (t^3/U^2) (a^2/\Phi_0)$, where $a$ is the lattice spacing and $\Phi_0$ - the elementary flux quantum.\cite{MotrunichChir} Thus, in this case a physical orbital magnetic field directly plays the analog of an external magnetic field, coupling to the emergent magnetic field $\nabla \times \vec{a}$ and, thereby, to the spinons, albeit with a reduced strength.

Like ordinary superconductors, the paired spin liquid/paired quantum Hall  phases exhibit two length scales: $\xi$ and $\lambda$, characteristic of fermion (spinon/neutral fermion) excitations and gauge field fluctuations, respectively. In the vicinity of the de-pairing transition, the behavior of these length scales is controlled by the RG fixed point describing the transition. Our scaling theory indicates that fermions disperse as $\omega \sim (|\vec{k}| - k_F)^{z_f}$ and gauge fluctuations disperse as $\omega \sim q^{z_b}$ with $z_b = 2 z_f = 2 + \epsilon$. 
As noted above, this relation holds both in the gapless spinon FS/HLR phase and at the pairing transition. Upon entering the paired phase, a characteristic energy scale $\omega = \Delta$ - the fermion gap is generated, which gives  $\xi \sim \lambda^2 \sim \Delta^{-1/z_f}$. So, as one approaches the transition, $\Delta \to 0$, and both the correlation length $\xi$ and the ``penetration depth" $\lambda$ diverge, albeit with different exponents. In particular, $\xi \gg \lambda$, so the paired phase in the vicinity of the transition is in the type-I regime, as was argued on general grounds in Ref.~\onlinecite{Sid1}. Further, the relation $E_v \gg \Delta$ obtained above is again typical of a type-I superconductor. Note that for $\epsilon = 1$, $\lambda \sim \Delta^{-1/3}$, which is the standard expression for the scaling of the physical (non-local Pippard) penetration depth in a conventional type-I superconductor.\cite{FetterWalecka}

As discussed above, the most dramatic manifestation of the type-I/type-II distinction in an ordinary superconductor is the response to an external magnetic field $H$. There is also an analog of this phenomenon for paired spinon/composite fermion phases.\cite{Sid1,Sid2} Let us begin with the QH case and first consider short-range electron-electron interactions.  In this case  the analog of the external magnetic field $H$ is the electron chemical potential $\mu$. The paired QH phase is incompressible, so for deviations of chemical potential $|\mu|$ smaller than some critical value, the system does not respond. (This is the analog of the Meissner effect in the superconductor). However, above a critical $\mu$, the electron density starts to change. This can occur in two ways: i) once $\mu > \mu_{c1}= 4 E_v$, charge $e/4$ quasiparticles (gauge field vortices) start to be nucleated. If the interactions between these quasiparticles are repulsive, we expect a stable dilute quasiparticle lattice to form. The Hall plateaux then persists for $\mu > \mu_{c1}$, as well as when one sweeps the physical magnetic field away from half-filling (holding the electron density constant). This QH counterpart of type-II superconducting behavior is thought to be realized in most conventional QH experiments.  ii) It is possible that the charge $e/4$ quasiparticles attract rather than repel, making the vortex lattice unstable. We then expect a first order phase transition between the paired QH phase and the HLR phase to occur at $\mu = \mu_c < 4 E_v$, accompanied by a jump in the electron density. This is the QH analogue of type-I superconducting behavior. We now show that this type-I scenario is, indeed, realized by paired QH states in the vicinity of the de-pairing transition. We can estimate the ``thermodynamic" critical chemical potential $\mu_c$, by equating the energy-densities of the paired state and the HLR state: $\epsilon_p = \epsilon_n - \frac{1}{2} \kappa \mu^2_c$, where $\kappa$ is the compressibility of the HLR phase. (We are measuring the chemical potential relative to the chemical potential of the HLR state at half-filling). Recalling our estimate, $\epsilon_n - \epsilon_p \sim \Delta^{1 + 1/z_f}$, we conclude $\mu_c \sim \Delta^{1/2 + 1/(2z_f)} = \Delta^{5/6} \ll E_v \sim \Delta^{1/3}$. Thus, the first order transition to the HLR phase occurs before individual $e/4$ quasiparticles can be excited, so the system is in the type-I regime. The magnitude of the density jump across the first-order transition is $\delta n_c = \kappa \mu_c \sim \Delta^{5/6}$ (see footnote \onlinecite{noteLL} for some caveats). For short-range electron-electron interactions, if one sweeps the magnetic field (holding the electron density fixed) away from $\nu = 1/2$, the system phase separates into macroscopic domains of the paired CF phase and the HLR phase. In practice, however, the first order transition between paired quantum Hall and the HLR phases will be rendered second order by the effect of impurities. Nevertheless, it is conceptually important to understand the nature of the transition in the clean limit.
 Long-range electron-electron interactions $U(\vec{x}) \sim 1/|\vec{x}|^{1+\epsilon}$ with $0 \leq \epsilon \leq 2$ frustrate the macroscopic phase separation, so one expect the formation of ``micro-emulsion"-like bubble/stripe phases in the vicinity of $\nu =1/2$.\cite{Sid1} 

A similar phenomenon can also occur in the $\mathbb{Z}_2$ spin-liquid phase in the vicinity of the de-pairing transition to the spinon FS phase.  Now, repeating the arguments presented above for the QH case, we expect an application of an external orbital magnetic field to induce a first order transition from the $\mathbb{Z}_2$ spin-liquid phase to the spinon FS phase at $\beta H_c \sim \Delta^{5/6}$, accompanied by a jump of magntidue $\sim \Delta^{5/6}$ in the spin-chirality. For spin-singlet pairing of spinons, the critical orbital field $H_c$ should be compared to the critical Zeeman field $H_Z = \Delta/(g_e \mu_B)$ needed to break up the Cooper pairs. In the strict $\Delta \to 0$, $H_Z$ is parametrically smaller than $H_c$, so the orbital effects may be neglected. This trend is further enhanced by the suppression of the coupling constant $\beta$ in the insulating regime $t\ll U$.

\section{Discussion}
We briefly discuss in sections \ref{sec:discIsing} - \ref{sec:discQH} a number of experimental phenomena to which our work is pertinent. In section \ref{sec:discnem}, we briefly note another system for which our RG results are relevant: the nematic phase in the continuum.

\subsection{Superconductivity near quantum critical points} 
\label{sec:discIsing}
One of the main results of this paper is a controlled theory of the superconducting instability of a quantum critical metal.  As an illustration we studied the Ising-nematic quantum critical point. Many of our results are expected to carry over to metals near other Pomeranchuk transitions. One of our main conclusions is that superconductivity is strongly enhanced near such 
quantum critical points. This gives  a firm theoretical basis to the empirical observation of superconducting domes with $T_c$ optimized near some putative quantum critical points.  The results on the Ising-nematic transition should be contrasted with those on the Mott transition from the spinon Fermi-surface spin-liquid insulator to a Fermi-liquid. We find that at such a Mott transition the pairing instability is suppressed. The Mott transition belongs to a qualitatively different class of QCPs in metals, one where an entire sheet of the electronic Fermi surface disappears through a continuous transition,\cite{ffl,kbpepinetal,mottcrit,mottcrit3d} and so displays a very different behavior compared to ``conventional" symmetry-breaking transitions.

 Returning to symmetry-breaking transitions, we recall that the problem of superconductivity near the spin-density-wave quantum critical point was addressed in 
Refs.~\onlinecite{maxsdw,BergQMC}. It was found there that non-Fermi liquid effects in the electron spectrum and pairing corrections
arose at similar energy scales, which preempted identification of a clear-cut non-Fermi liquid regime in the normal state.

For the specific case of the Ising-nematic transition, experiments\cite{pnic6} on electron-doped iron superconductor Ba(Fe$_{1-x}$Co$_x$)$_2$As$_2$ show that a quantum critical point associated with such order likely lies directly underneath the superconducting dome. This quantum critical point appears to be separated from a different one associated with onset of spin density wave order that occurs at lower $x$. It is tempting to attribute the optimality of the superconducting $T_c$ to the enhanced fluctuations of the underlying Ising-nematic quantum critical point. For this scenario to be legitimate it is necessary that other fluctuations (for instance in the spin density) have much weaker effects in the normal state  at optimal doping. A contrasting scenario likely applies to a different iron superconductor BaFe$_2$(As$_{1-x}$P$_x$)$_2$ obtained by isovalent substitution.\cite{MatsudaReview} In this case, optimal $T_c$ occurs around $x = 0.30$, which is roughly where the Neel temperature associated with spin density wave (SDW) order (present at low $x$) extrapolates to zero.  The strong enhancement of the NMR relaxation rate near the optimal doping further suggests the presence of an SDW critical point.\cite{FeAsNMR} The low-$x$ material also displays Ising-nematic order but, according to some reports,\cite{pnic7} it disappears at a larger $x$ that is near the {\it overdoped} edge of the superconducting dome. So the SDW fluctuations seem to dominate over any nematic fluctuations near optimal doping in this material.

Quantum critical nematic fluctuations may also play a role in the physics of nearly optimally doped cuprates, and our results may provide a foundation to assessing their effects.

A different aspect of our theoretical results is the relationship between non-Fermi liquid physics and superconductivity near the Ising-nematic QCP. In the small-$\epsilon$ regime where our calculations are controlled we found that the superconductivity is so strong that it sets in at a temperature scale parametrically larger than the scale at which non-Fermi liquid effects set in. For $\epsilon =1$ we expect that there is no such separation and the two phenomena happen at parametrically the same scale. In this case, it is possible that the superconductivity will rear its head before the non-Fermi liquid physics has fully developed. It is then interesting to ask what happens when the superconductivity is suppressed with an external magnetic field. Presumably, this will expose the non-Fermi liquid physics of the Ising-nematic quantum critical point down to low temperatures. In particular, the specific heat will follow the predicted power law $C \sim T^{\frac{2}{3}}$. 
Some aspects of the non-Fermi liquid physics predicted for the critical point will likely be suppressed by the magnetic field, along with the superconductivity. A good example is the low-energy tunneling density of states, which in the absence of the magnetic field was found to be power-law suppressed at the QCP\cite{MS,Mross}. This effect arises primarily from enhanced Cooper pairing fluctuations\cite{Mross}. Since these will be suppressed in a magnetic field, so will the singularity in the tunneling density of states.

\subsection{When is superconductivity enhanced near a quantum critical point?}     
\label{sec:discHF}
Based on our results we now suggest an answer to the empirical puzzle that superconductivity is enhanced at some, but not all, metallic quantum critical points. As we emphasized in the introduction, there are actually two qualitatively distinct classes of such quantum critical points distinguished primarily by the fate of the electron Fermi surface. For criticality associated primarily with the onset of broken symmetry order, the electron Fermi surface evolves continuously but is distorted by the broken symmetry. On the other hand, there are continuous quantum phase transitions where the electron Fermi surface evolves discontinuously. The associated quantum critical points are dominated by fluctuations of the electronic structure itself.  

The specific examples studied in this paper lead us to suggest more generally  that quantum critical points associated primarily with the onset of broken symmetry will show enhanced superconductivity, while those associated primarily with a discontinuous jump of the Fermi surface may not.  Apart from the specific results in this paper, this suggestion finds theoretical support in many previous approximate treatments of superconductivity  due to order parameter fluctuations, {\it e.g.} at the SDW onset QCP. There are currently very few theoretical examples of transitions in the second class (with a disappearing  Fermi surface). One known example is given by the Mott transition from a spinon FS spin-liquid insulator to a Fermi-liquid; here we showed that pairing is suppressed at the QCP. Whether this is a general property of the second class of transitions is a good target for future research. As we noted in section \ref{sec:introMott}, the lack of long-range superconducivity at a Mott transition is not surprising; a far more non-trivial conclusion of our analysis is the suppression of local pairing correlations near the transition, reflected in the vanishing of the single electron gap at the QCP, and the absence of superconductivity in the Fermi-liquid phase adjacent to the QCP.


Our suggestion finds empirical support in the absence of superconductivity in the heavy fermion materials CeCu$_{6-x}$Au$_x$ and YbRh$_2$Si$_2$ near their quantum critical points. 
These have long been thought\cite{LohneysenReview} to be systems where the Fermi surface evolves discontinuously and the electronic structure changes due to the breakdown of Kondo screening.  
A futher complication in these materials is that the critical point seems to involve fundamental changes of the electronic structure accompanied also by the onset of antiferromagnetic order. There is currently no theoretical understanding of why the discontinuous Fermi surface change coincides with the onset of broken symmetry.   Nevertheless, it is natural to expect that the fate of superconductivity will then be determined  by
a delicate interplay between two competing effects: the pairing tendencies of order parameter fluctuations  and the strong non-Fermi liquid effects due to the electronic fluctuations. 
We hope that the present paper sets the stage for future progress on this issue. 

\subsection{Quantum Hall states at $\nu = 1/2$.} 
\label{sec:discQH}
In this paper, we have developed a systematic theory of the transition between the compressible HLR phase and the incompressible Moore-Read phase of the quantum Hall fluid. We have demonstrated that contrary to previous theoretical claims\cite{BonesteelPairBreak} a direct continuous transition between these phases is allowed. 
In a conventional GaAs system, the HLR phase is believed to be realized at filling factors $\nu = 1/2$, $\nu = 3/2$, while the Moore-Read phase is a candidate for the plateaux at $\nu  = 5/2$. In a large magnetic field, when the mixing between Landau levels can be neglected, the physics of a partially filled Landau level is determined by the projection of the electron-electron interaction onto the states in the Landau level. The projection is different for different Landau levels, which is believed to explain the above contrasting behaviors observed in the lowest ($n =0$) and first ($n = 1$) Landau levels.\cite{Morf, HaldaneRezayi} Since both the electron-electron interactions and the form of the single-particle states in GaAs are difficult to tune, the realization of a direct transition between the Moore-Read phase and the HLR phase in GaAs is challenging. However, it was recently suggested that this transition may be realized in bilayer graphene, by tuning the strength of the perpendicular electric field.\cite{Apalkov, Zlatko} The introduction of the perpendicular electric field modifies the form of the single-particle states forming the Landau level and hence the effective interactions within the Landau level. We, thus, hope that bilayer graphene will provide an experimental avenue to test our predictions.

\subsection{Nematic phase in the continuum}
\label{sec:discnem}
In this paper, we have concentrated on the Pomeranchuk instability in the presence of a lattice, where a discrete rotational symmetry is spontaneously broken. It is also interesting to consider a system in the continuum, which possesses a full SO(2) rotational symmetry. In this case, the nematic phase where SO(2) is spontaneously broken to a discrete two-fold subgroup will possess a gapless Goldstone mode. A curious feature of this phase is that unlike in systems involving spontaneous breaking of internal (non-spatial) symmetries, here the Goldstone mode couples to electrons near the FS in a non-derivative manner.\cite{OKF01} As a result, the entire nematic phase was predicted to exhibit non-Fermi-liquid behavior.\cite{OKF01} In fact, it is easy to see, that the coupling of the Goldstone mode to the FS is essentially the same as for the gapless critical mode $\phi$ at the Ising-nematic QCP.  
Therefore, the effective theory of the  ``continuum"-nematic  phase will be identical to that of the Ising-nematic QCP studied above. In particular, according to the results of our $\epsilon$-expansion, the entire continuum-nematic phase will become superconducting and the non-Fermi-liquid behavior of Ref.~\onlinecite{OKF01} will be preempted.


{\it Acknowledgements.}
We would like to thank S.~Parameswaran, Z.~Papic, A.~Chubukov, S.-S.~Lee, P.~A.~Lee, D.~Scalapino, S.~Kivelson, C.~Nayak, L.~Balents and M.~P.~A.~Fisher for useful discussions. This research was supported in part by the National Science Foundation under Grant Nos. NSF PHY11-25915
and DMR-1360789,
and by the Templeton Foundation. 
The work of TS was supported by Department of Energy DESC-8739- ER46872, and partially by a Simons Investigator grant from the Simons Foundation. Research at Perimeter Institute is supported by the Government of Canada through Industry Canada 
and by the Province of Ontario through the Ministry of Research and Innovation.

\appendix
\section{Solution of RG equations: Ising-nematic QCP}
\label{sec:RGnem}
In this appendix we provide a detailed solution of the RG equations, (\ref{eq:RGalpha}), (\ref{eq:FlowV}) for the Ising-nematic QCP ($\zeta = 1$ in Eq.~(\ref{eq:FlowV})). We use the notation $\widetilde{\alpha} = \alpha/N$.

We begin by considering the case  $\epsilon = 0$. Then $\widetilde{\alpha}$ runs logarithmically to zero,
\beq \widetilde{\alpha}(\ell) = \frac{\widetilde{\alpha}(0)}{1 + \widetilde{\alpha}(0) \ell}, \quad v_F(\ell) = \frac{v_F(0)}{1+ \widetilde{\alpha}(0) \ell} \label{flowalpha}\eeq
and the non-Fermi liquid corrections become appreciable below an energy scale of order $e^{-\ell_{\mathrm{nFL}}}$ with  $\ell_{\mathrm{nFL}} \sim 1/\widetilde{\alpha}$. 

Since the flow of $\widetilde{\alpha}$ is slow, in order to analyze the flow of $\widetilde{V}$, let us first assume that $\widetilde{\alpha}$ is constant. We see that the flow is then towards $\widetilde{V} = -\infty$ signaling a pairing instability. Solving Eq.~(\ref{eq:FlowV}),
\beq \widetilde{V}(\ell) = \sqrt{\widetilde{\alpha}}\tan\left(- \sqrt{\widetilde{\alpha}} \ell + \tan^{-1}\frac{\widetilde{V}(0)}{\sqrt{\widetilde{\alpha}}}\right) \label{flowV1}\eeq
We see that $\widetilde{V}$ diverges at
\beq \ell_{p} = \frac{1}{\sqrt{\widetilde{\alpha}}} \left(\frac{\pi}{2} + \tan^{-1}\frac{\widetilde{V}(0)}{\sqrt{\widetilde{\alpha}}}\right) \label{app:ell1}\eeq
and we expect a pairing gap $\Delta \sim \Lambda_\omega e^{-\ell_{p}}$. Let us discuss various limits of Eq.~(\ref{app:ell1}). If the ``bare" short range interaction is small compared to the long-range interaction, 
\beq \ell_{p} \approx \frac{\pi}{2 \sqrt{\widetilde{\alpha}}}, \quad |\widetilde{V}| \ll \sqrt{\widetilde{\alpha}} \label{Vweak}\eeq
On the other hand, if the bare short range interaction is large and repulsive, 
\beq \ell_{p} \approx \frac{\pi}{\sqrt{\widetilde{\alpha}}}, \quad \widetilde{V} \gg \sqrt{\widetilde{\alpha}}\label{Vstrongpos}\eeq
{\em i.e.\/} the magnitude of the gap is reduced by a factor of two on the logarithmic scale compared to the case (\ref{Vweak}),
Finally, if the bare short range interaction is large and attractive,
\beq \ell_{p} \approx \frac{1}{|\widetilde{V}|}, \quad \widetilde{V}< 0,\,  |\widetilde{V}| \gg  \sqrt{\widetilde{\alpha}} \label{Vstrongneg} \eeq
which is just the usual BCS result. In any case, $\widetilde{\alpha} \ell_{p} <  \pi \sqrt{\widetilde{\alpha}} \ll 1$ hence the running of $\widetilde{\alpha}$ can, indeed, be neglected in estimating the size of the gap. Moreover, $\ell_{p} < \pi/ \sqrt{\widetilde{\alpha}} \ll \ell_{\mathrm{nFL}} \sim 1/\widetilde{\alpha}$, hence the non-Fermi-liquid physics is preempted by pairing.

\begin{figure}[t]
\begin{flushleft}
\includegraphics[width=3.5in]{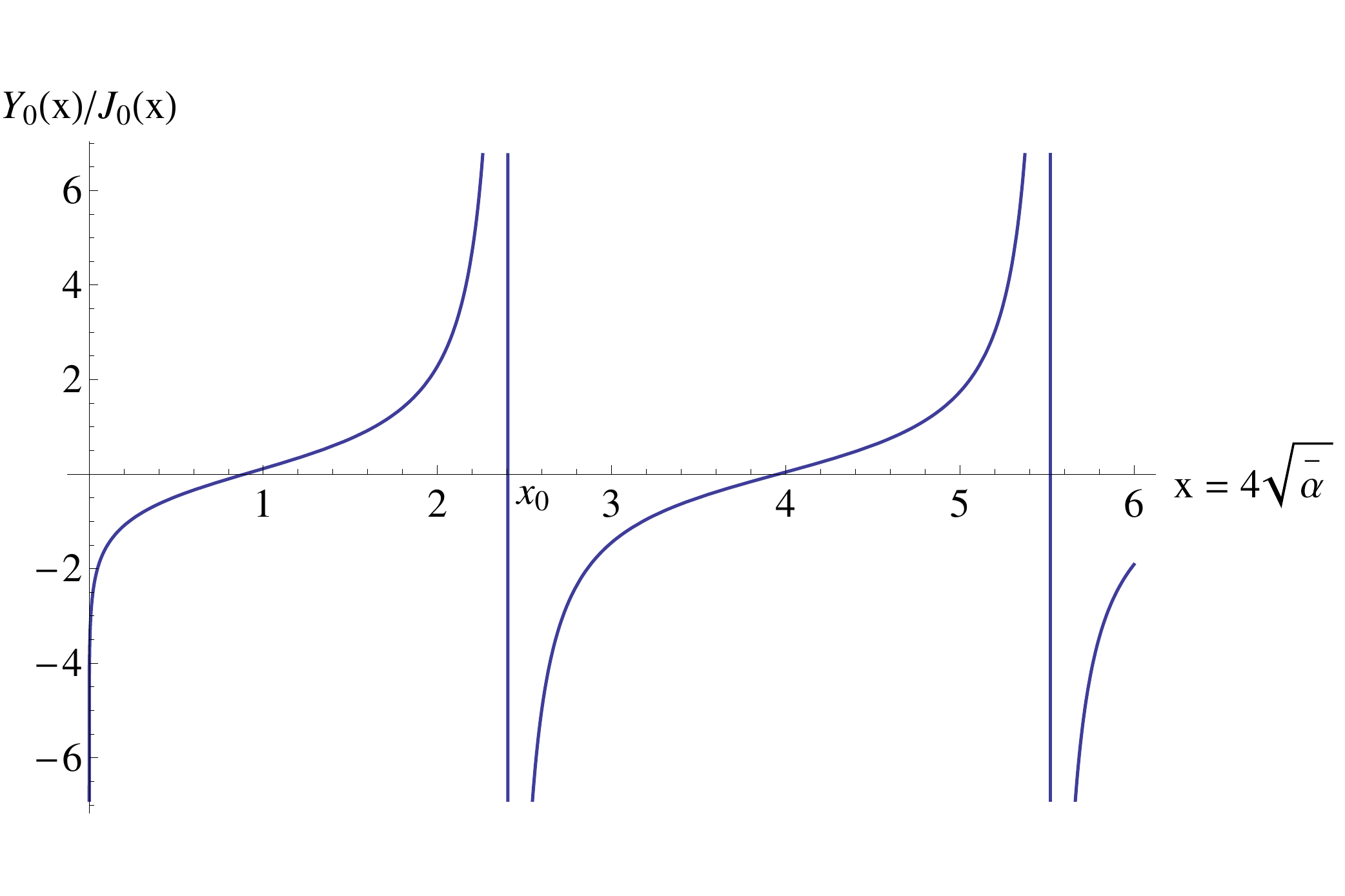}
\caption{Determination of the pairing scale $\ell_{p}$ and the corresponding value $\bar{\alpha}_{p}$ in the regimes $\widetilde{\alpha} \sim O(\epsilon^2)$ and  $\widetilde{\alpha} \ll O(\epsilon^2)$ (see Eqs.~(\ref{alphapair}), (\ref{alphapair2})).}
\label{FigBess}
\end{flushleft}
\end{figure}

If we turn on a finite $\epsilon$, the flow of $\widetilde{V}$ to $-\infty$ persists. Let us estimate the size of the gap. In the present case $\widetilde{\alpha}$ and $v_F$ run as,
\bea \widetilde{\alpha}(\ell) &=& \frac{\widetilde{\alpha}(0) e^{\epsilon \ell/2}}{1 + \displaystyle \frac{2 \widetilde{\alpha}(0)}{\epsilon} (e^{\epsilon \ell/2} - 1)} \nn \\ v_F(\ell) &=& \frac{v_F(0) }{1 + \displaystyle \frac{2 \widetilde{\alpha}(0)}{\epsilon} (e^{\epsilon \ell/2} - 1)}. \eea
Note that for $\widetilde{\alpha} \sim \widetilde{\alpha}_* = \epsilon/2$ the scale at which non-Fermi-liquid effects become manifest is $\ell_{\mathrm{nFL}} \sim 1/\epsilon$. For $\widetilde{\alpha} \gg \epsilon$, one first observes logarithmic running of $v_F$ for $\ell \gtrsim \ell_{\mathrm{nFL}} \sim 1/\widetilde{\alpha}$, and then power law running for $\ell \gtrsim 1/\epsilon$. Finally, if $\widetilde{\alpha} \ll \epsilon$, $\ell_{\mathrm{nFL}} \sim \frac{2}{\epsilon} \log(\epsilon/\widetilde{\alpha})$. Proceeding to the flow of $\widetilde{V}$, we observe that if $\widetilde{\alpha} \gg \epsilon^2$, the flow of $\widetilde{\alpha}$ can be ignored for the purpose of estimating the pairing scale and previous results Eqs.~(\ref{flowV1}) and (\ref{app:ell1}) hold. Comparing the pairing scale and the non-Fermi-liquid scale, we find that the former is always parametrically higher in this regime. Indeed, for $\widetilde{\alpha} \sim O(\epsilon)$ and $\widetilde{\alpha} \gg \epsilon$, $\ell_{p} < \pi/\sqrt{\widetilde{\alpha}} \ll \ell_{\mathrm{nFL}} \sim 1/\widetilde{\alpha}$, while for $\epsilon^2 \ll \widetilde{\alpha} \ll \epsilon$, $\ell_{p} < \pi/\sqrt{\widetilde{\alpha}} \ll \ell_{\mathrm{nFL}} \sim \frac{2}{\epsilon} \log \frac{\epsilon}{\widetilde{\alpha}}$. In the remaining regime $\widetilde{\alpha} \lesssim \epsilon^2$, the flow of $\widetilde{\alpha}$ cannot be ignored. However, this regime can be effectively analyzed as a part of a wider range $\widetilde{\alpha} \ll \epsilon$. As is already clear from the arguments above, if we start with $\widetilde{\alpha} \ll \epsilon$, $\widetilde{\alpha}$ will remain in this range throughout the evolution. Hence, in this regime, we may approximate,
\beq \frac{d \widetilde{\alpha}}{d\ell} = \frac{\epsilon}{2} \widetilde{\alpha}\eeq
\beq \widetilde{\alpha} \frac{d \widetilde{V}}{d \widetilde{\alpha}} = -\frac{2}{\epsilon}(\widetilde{\alpha} + \widetilde{V}^2) \label{dVdalpha}\eeq
We may eliminate the $\epsilon$ dependence from Eq.~(\ref{dVdalpha}) by defining $\widetilde{\alpha} = \epsilon^2 \bar{\alpha}$, $\widetilde{V} = \epsilon \bar{V}$. Then,
\beq \bar{\alpha}\frac{d \bar{V}}{d \bar{\alpha}} = - 2 (\bar{\alpha} + \bar{V}^2)\label{dVdalphabar}\eeq
The solution to Eq.~(\ref{dVdalphabar}) is,
\beq \bar{V}(\bar{\alpha}) = -\frac{\sqrt{\bar{\alpha}} (J_1(4\sqrt{\bar{\alpha}}) + C Y_1(4 \sqrt{\bar{\alpha}}))}{J_0(4 \sqrt{\bar{\alpha}}) + C Y_0(4 \sqrt{\bar{\alpha}})}\eeq
where initial conditions fix the constant $C$ to be,
\beq C = -  \frac{\sqrt{\bar{\alpha}(0)} J_1(4 \sqrt{\bar{\alpha}(0)}) + J_0(4 \sqrt{\bar{\alpha}(0)}) \bar{V}(0)}{\sqrt{\bar{\alpha}(0)} Y_1(4 \sqrt{\bar{\alpha}(0)}) + Y_0(4 \sqrt{\bar{\alpha}(0)}) \bar{V}(0)}\eeq
Observe that $\bar{V}$ has a divergence at $\bar{\alpha} = \bar{\alpha}_p$ with 
\beq \frac{J_0(4 \sqrt{\bar{\alpha}_p})}{Y_0(4\sqrt{\bar{\alpha}_p})} = - C \label{alphapair}\eeq
As is clear from Fig.~\ref{FigBess}, irrespective of the value of $C$, if $\alpha$ ($\bar{\alpha}$) is of $O(\epsilon^2)$ (of $O(1)$) or less, the above equation has a solution with $\bar{\alpha}_{p}$ at most of $O(1)$. Hence,
\beq \ell_{p} = \frac{2}{\epsilon} \log \frac{\epsilon^2 \bar{\alpha}_{p}}{\widetilde{\alpha}} \ll \ell_{\mathrm{nFL}} \sim \frac{2}{\epsilon} \log\frac{\epsilon}{\widetilde{\alpha}} \label{ellpair}\eeq 
Thus, in this regime pairing also always occurs before non-Fermi-liquid effects become significant. Having established this, we will not analyze the full behaviour of the pairing scale as a function of initial $\widetilde{\alpha}$ and $\widetilde{V}$ in this regime, but will only discuss the case of smallest coupling, $\widetilde{\alpha} \ll \epsilon^2$ ($\bar{\alpha} \ll 1$). Then, Eq.~(\ref{alphapair}) may be rewritten as,
\beq \frac{Y_0(4 \sqrt{\bar{\alpha}_{p}})}{J_0(4\sqrt{\bar{\alpha}_{p}})} \approx \frac{Y_0(4 \sqrt{\bar{\alpha}})}{J_0(4\sqrt{\bar{\alpha}})} - \frac{1}{2 \pi \bar{V}} \label{alphapair2}\eeq

The function $Y_0(x)/J_0(x)$ is increasing wherever it is continuous (see Fig.~\ref{FigBess}). Hence, as we need a solution with $\bar{\alpha}_{p} > \bar{\alpha}$, for $\bar{V} > 0$ we conclude that $4 \sqrt{\bar{\alpha}}_{p} > x_0$, with $x_0 \approx 2.405 $ - the first zero of $J_0(x)$. Moreover, as $Y_0(x) \approx \frac{2}{\pi} \log x$ for $x \to 0$, the right-hand-side of Eq.~(\ref{alphapair2}) tends to $-\infty$, hence,
\beq \bar{\alpha}_{p} \approx \frac{x^2_0}{16} \approx 0.361, \quad \bar{V} > 0,\, \widetilde{\alpha} \ll \epsilon^2, \label{alphapair3} \eeq
and the pairing scale can be obtained from Eq.~(\ref{ellpair}).  Now, if $\bar{V} < 0$ but $|\bar{V}| \ll (\log \frac{1}{\bar{\alpha}})^{-1}$, the right-hand-side of Eq.~(\ref{alphapair2}) tends to $+ \infty$ and Eq.~(\ref{alphapair3}) still holds. Finally, if $\bar{V} < 0$ and $|\bar{V}| \gg (\log \frac{1}{\bar{\alpha}})^{-1}$, $\bar{\alpha}_{p} \ll 1$ and from Eq.~(\ref{alphapair2}) we obtain
\beq \bar{\alpha}_{p} = \bar{\alpha} e^{1/2|\bar{V}|}, \quad \bar{V} < 0,\, |\bar{V}| \gg \left(\log \frac{1}{\bar{\alpha}}\right)^{-1},\, \widetilde{\alpha} \ll \epsilon^2 \eeq
which gives the standard BCS result $\ell_{p} = \frac{1}{|\widetilde{V}|}$.

\section{Solution of RG equations: HLR phase with Coulomb interactions}
\label{sec:RGHLR}
In this appendix we provide a detailed solution of the RG equations, (\ref{eq:RGalpha}), (\ref{eq:FlowV}), for the HLR phase with Coulomb interactions. We, thus, set $\zeta = -1$ in Eq.~(\ref{eq:FlowV}) and $\epsilon = 0$ in Eq.~(\ref{eq:RGalpha}). 

The flow in the $(\widetilde{V},\tilde{\alpha})$ plane takes the form shown in Fig.~\ref{fig:Vflow2}. Note that part of the phase space is controlled by the fixed point $\widetilde{V} = 0$, $\tilde{\alpha} = 0$, while the rest of the flow trajectories are towards $\widetilde{V} = -\infty$. We will show below that the two regions are separated by the line $\widetilde{V} = -\sqrt{\widetilde{\alpha}}$. 

Let us proceed to solve the flow equations. The flow of $\widetilde{\alpha}$ is the same as for the Ising-nematic case, Eq.~(\ref{flowalpha}). Defining $\widetilde{V} = \sqrt{\widetilde{\alpha}} g$,
\beq \frac{dg}{d \ell} = \sqrt{\widetilde{\alpha}}(1-g^2) + \frac{1}{2} \widetilde{\alpha} g \label{dfdl}\eeq
In the limit $\widetilde{\alpha} \to 0$, we can neglect the last term in Eq.~(\ref{dfdl}). Then,
\beq \frac{dg}{d \widetilde{\alpha}} = - \widetilde{\alpha}^{-3/2} (1-g^2)\eeq
which gives,
\beq g(\ell) = \frac{(g(0)+1) \exp\left[4(\widetilde{\alpha}(\ell)^{-1/2} - \widetilde{\alpha}(0)^{-1/2})\right] + (g(0) - 1)}{(g(0)+1) \exp\left[4(\widetilde{\alpha}(\ell)^{-1/2} - \widetilde{\alpha}(0)^{-1/2})\right] - (g(0) - 1)} \label{fsol}\eeq
As $\widetilde{\alpha}$ flows to zero the following cases are possible. If $g(0) > - 1$, {\em i.e.\/} $\widetilde{V}(0) > -\sqrt{\widetilde{\alpha}(0)}$, then $g$ flows to $1$, 
\beq \widetilde{V}(\ell) \to \sqrt{\widetilde{\alpha}(\ell)} \approx \frac{1}{\sqrt{\ell}},\quad  \widetilde{V} > -\sqrt{\widetilde{\alpha}}\eeq
If one starts exactly on the transition line, $g(0) = -1$, then $g$ remains fixed at $g = -1$, 
\beq \widetilde{V}(\ell) = - \sqrt{\widetilde{\alpha}(\ell)} \approx -\frac{1}{\sqrt{\ell}},\quad \widetilde{V} = -\sqrt{\widetilde{\alpha}(0)}\eeq
Finally, if $g(0) < -1$, {\em i.e.\/} $\widetilde{V}(0) < -\sqrt{\widetilde{\alpha}(0)}$, $\widetilde{V}(\ell)$ flows to $-\infty$ and diverges at $\ell = \ell_{p}$, with
\beq \ell_{p} = \frac{1}{16} \left(\log \frac{\widetilde{V} - \sqrt{\widetilde{\alpha}}}{\widetilde{V} + \sqrt{\widetilde{\alpha}}} + 4 \widetilde{\alpha}^{-1/2}\right)^2 - \widetilde{\alpha}^{-1}\eeq
In particular, as $\widetilde{V}$ approaches the transition line, $\delta V = \widetilde{V} + \sqrt{\widetilde{\alpha}} \to 0^-$, the pairing gap $\Delta \sim \Lambda_\omega e^{-\ell_{p}}$ vanishes as,
\beq \Delta \sim \Lambda_\omega \exp\left(-\frac{1}{16} \log^2 |\delta V|\right) = |\delta V|^{\frac{1}{16} \log\frac{1}{|\delta V|}}\label{Deltalog2}\eeq


\begin{thebibliography}{}
\bibitem{NematicFO}  While an analysis of the effective field theory governing the Ising-nematic QCP shows
that fluctuations do not drive the transition first order, many microscopic models believed to harbor such a QCP
display a first order transition already at the mean field level.\cite{KKC03, YOM05} However, with band-structure and interaction
parameters in the appropriate region, the first order transition can be avoided in the mean field treatment.\cite{YOM05}
Moreover, even when the transition is first order at the mean field level, fluctuations may drive it second order as found in
Ref.~\onlinecite{MetznerFRG}.
\bibitem{FNAmperian} In this paper, we only consider conventional BCS pairing of spinons with total pair momentum $\vec{P} = 0$. A different possibility, encouraged by the Amperian current-current interaction, where the total pair momentum $|\vec{P}|= 2 k_F$ was considered in Ref.~\onlinecite{Ampere}. However, in the $\epsilon$-expansion employed in the present paper, the spinon FS is stable to such pairing.
\bibitem{Loop} Strictly speaking, it is not known whether higher loop terms in the RPA resummed expansion are suppressed by powers of $\alpha$. Thus, the validity of the perturbative expansion in $\epsilon$ with $N$ - finite, as well as of the results at $\epsilon = 0$ both with $N$ finite and with $N \to \infty$, has not been firmly established. On the other hand, the calculations in the double-scaling limit $\epsilon \to 0$, $N \to \infty$, $N \epsilon \sim O(1)$ are certainly controlled.\cite{Mross} In the present paper, we will assume that the first expansion is controlled, as well.
\bibitem{FNcircular} Strictly speaking, this assumption is unphysical for the Ising-nematic QCP, however, it is not expected to change most of our results qualitatively (see end of section \ref{sec:NematicRes} for further discussion).
\bibitem{noteLL} Strictly speaking, the HLR phase is only defined at zero temperature, where it occurs at a fixed electron density ($\nu  =1/2$ for the clean, Gallilean invariant system). For $\nu$ deviating from $1/2$ and $T = 0$, composite fermions see a background magnetic field $B_{\mathrm{eff}} = B - 2 (2\pi) n$, and so form Landau levels. At low enough temperature, one, thus, expects to see a sequence of integer QH states of composite fermions, which correspond to a sequence of FQH states of physical electrons with $\nu = p/(2p \pm 1)$.\cite{HLR} In our discusion, we assume that the temperature $T$ is much larger than the cyclotron frequency $\omega_c$ of composite fermions, so that  Landau-level quantization can be neglected. At the same time, for our estimates to be correct, we need the temperature to be much lower than the gap, $\Delta$, of the paired CF phase. We expect  the cyclotron frequency, $\omega_c = \frac{B_{\mathrm{eff}}}{m^*} = \frac{v_F B_{\mathrm{eff}}}{k_F}$. The Fermi-velocity in the HLR phase runs with energy, (\ref{eq:Floweps}), $v_F \sim \omega^{1-1/z_f}$. When Landau-levels of composite fermions form, we expect that the running of $v_F$ will be cut-off at $\omega = \omega_c$. We, thus, obtain, $\omega_c \sim (B_{\mathrm{eff}})^{z_f}$. At the first order transition from the paired CF phase to the HLR phase, $B_{\mathrm{eff}} \sim \delta n_c \sim \Delta^{5/6}$, so $\omega_c \sim \Delta^{5/4}$. Thus, our calculations are legitimate as long as $\Delta^{5/4} \ll T \ll \Delta$.
\end{thebibliography}
\end{document}